\documentclass{sig-alternate-no-permission-block-space} 
\usepackage{cite,subfigure}
\usepackage{amsfonts}
\usepackage{amsmath}
\usepackage{url}
\usepackage{tikz,algonjk}
\usepackage{xspace}
\usepackage{mdwlist}
\usepackage{hyperref}
\hypersetup{
  pdftitle={Slick Packets},
  pdfauthor={Giang T. K. Nguyen, Rachit Agarwal, Junda Liu, Matthew Caesar, P. Brighten Godfrey, Scott Shenker},
}

\makeatletter
\def\blfootnote{\xdef\@thefnmark{}\@footnotetext}
\makeatother
\newtheorem{theorem}{Theorem}
\newcommand{\paragraphb}[1]{\vspace{0.03in}\noindent{\bf #1.} }

\newcommand{\FS}{\textrm{FS}\xspace}
\usepackage{anysize}
\marginsize{2cm}{2cm}{1cm}{3cm}

\newcommand{\name}{{\sc Slick\-Pac\-kets}\xspace}
\newcommand{\nametitle}{SLICKPACKETS\xspace}
\newcommand{\deto}{{alternate}\xspace}
\newcommand{\detosingular}{{an alternate}\xspace}
\newcommand{\sg}{SafeGuard\xspace}
\newcommand{\dist}{{\operatorname{dist}}}
\newcommand{\hopcount}{{\operatorname{h}}}
\newcommand{\codecA}{{Default}\xspace}
\newcommand{\codecB}{{Direct}\xspace}

\title{Slick Packets}
\numberofauthors{6} 
\author{
\alignauthor
Giang~T.~K.~Nguyen\\
       \affaddr{University of Illinois}\\
       \affaddr{at Urbana-Champaign, USA}\\
       \email{nguyen59@illinois.edu}
\alignauthor
Rachit Agarwal\\
\affaddr{University of Illinois}\\
\affaddr{at Urbana-Champaign, USA}\\
\email{agarwa16@illinois.edu}
\alignauthor Junda Liu\\
\affaddr{University of California}\\
\affaddr{at Berkeley, USA}\\
\email{liujd@cs.berkeley.edu}
\and  
\alignauthor Matthew Caesar\\
\affaddr{University of Illinois}\\
\affaddr{at Urbana-Champaign, USA}\\
\email{caesar@illinois.edu}
\alignauthor P. Brighten Godfrey\\
\affaddr{University of Illinois}\\
\affaddr{at Urbana-Champaign, USA}\\
\email{pbg@illinois.edu}
\alignauthor Scott Shenker\\
\affaddr{University of California}\\
\affaddr{at Berkeley, USA}\\
\email{shenker@cs.berkeley.edu}
}

\begin{document}

\clubpenalty=10000
\widowpenalty = 10000
\maketitle

\section*{Abstract}

\textnormal{Source-controlled routing has been proposed as a way to improve flexibility of future network architectures, as well as simplifying the data plane. However, if a packet specifies its path, this precludes fast local re-routing within the network. We propose \name, a novel solution that allows packets to slip around failures by specifying alternate paths in their headers, in the form of compactly-encoded directed acyclic graphs. We show that this can be accomplished with reasonably small packet headers for real network topologies, and results in responsiveness to failures that is competitive with past approaches that require much more state within the network. Our approach thus enables fast failure response while preserving the benefits of source-controlled routing\blfootnote{This is an extended version of a paper that appeared in ACM SIGMETRICS 2011. Supporting code is available at \url{http://code.google.com/p/slick-packets/}}.}


\section{Introduction}

Traditional routing protocols are {\bf network-controlled}: routes are computed within the network, with each router picking, from among its neighbors, the next-hop to each destination. Examples include BGP for interdomain routing, and OSPF for intradomain routing. An alternate paradigm, {\bf source-controlled routing} (SCR), improves the flexibility of the network architecture. Rather than computing all routes within the network, SCR architectures~\cite{xu06miro,yang06,nira,motiwala08,pathlets} reserve some choice of routes for the {\em sources}\footnote{In this paper, we use ``source'' to refer either to end-hosts or to edge routers acting on their behalf.} to select on a per-packet basis. The uses of SCR's routing flexibility are quite diverse. Sources can observe end-to-end reliability problems and switch to a working path within a few round-trip times (RTTs); pick better-performing routes based on observed performance~\cite{andersen01ron,savage99detour,gummadi04improving}; improve load balance since path selection is finer-grained~\cite{qiu03}; encourage competition among network pro\-viders~\cite{clark02tussle}; improve security~\cite{wa+06}; or optimize for other application-specific objectives. SCR is thus a promising approach to improve the flexibility of the network layer in future Internet architectures.

However, one remaining problem is that of {\em fast failure reaction}.  This problem arose in early network-controlled routing (NCR) protocols, which suffered from unreliability during network dynamics: during the distributed convergence process, packets could enter ``black holes'' or loops, resulting in tens of seconds or minutes of downtime in Internet end-to-end paths~\cite{icmbd02,1159956}.  Treating these basic protocols as a baseline, two high-level approaches have been proposed to improve failure reaction.

The first approach works within the NCR paradigm by computing \detosingular path to each destination (or IP prefix or AS); a router can locally switch to the \deto path without waiting for a control-plane convergence process.  Packets can thus be delivered continuously, except for the minimal time it takes for a router to detect failure of one of its directly connected links and locally switch to an alternate path.  Examples include MPLS Fast Reroute~\cite{rfc4090}, \sg~\cite{safeguard}, and FCP~\cite{fcp} for intradomain routing, and R-BGP~\cite{r-bgp} for interdomain routing.  However, this approach lacks the routing flexibility of SCR.

A second approach to improve failure reaction is to leverage SCR's routing flexibility: a source can switch routes without waiting for the Internet's control plane to reconverge. While this improves failure reaction time relative to the baseline above, the source still must wait to receive notice of the failure. Regardless of the means of notification, this will take at least on the order of one RTT, which at Internet scales would be much slower than the first approach of using NCR with \deto paths. And in the SCR proposals that provide the most flexibility~\cite{nira,pathlets}, sources specify in the packet header an explicit route (perhaps at the level of autonomous systems) rather than a destination, so the NCR and SCR techniques cannot be immediately combined.

The goal of this paper is to achieve the best of two worlds: the fast failure reaction of \deto routes embedded within the network, and the flexibility of routes chosen by sources at the edge of the network. To meet this goal, we work within the SCR paradigm, but with a twist.  Instead of specifying a single path to the destination, the packet header contains a directed acyclic graph that we call the {\em forwarding subgraph} (FS). Each router along the packet's path may choose to forward it along any of the outgoing links at that router's node in the FS (optionally preferring a path marked as the primary), with no danger of causing a forwarding loop. This approach, which we call \name, allows packets to ``slip'' around failures in-flight while retaining the flexibility of source route control.  Moreover, \name provides a scalability benefit over NCR with \deto paths: rather than requiring multiple routes to every destination in every router's forwarding table, \name routers need only local information.

Of course, our approach also presents several challenges.  Chief among these is how to encode an FS with sufficient path diversity into the small space afforded by a packet header.  We introduce techniques through which the FS can be encoded compactly enough for our mechanism to be feasible.  For example, an FS providing \detosingular path at every hop along the primary occupies less than 26 bytes for $99$\% of evaluated source-destination pairs in an AS-level Internet map, and no higher than 50 bytes in all evaluated cases.  Thus, the technique incurs manageable overhead for applications that send packets of moderate to large size.  We also demonstrate through a simulation-based performance evaluation that \name achieves failure reaction performance that is comparable to the best of NCR architectures~\cite{safeguard}.

The rest of this paper proceeds as follows. In \S\ref{sec-overview}, we present an overview of \name and its principal design challenges. \S\ref{sec-design} gives a detailed presentation of the \name design. We evaluate the performance of our design in terms of header size and failure reaction in \S\ref{sec-eval}.  We discuss extensions of \name in \S\ref{sec-extensions} and related work in \S\ref{sec-relatedwork}, and conclude in \S\ref{sec:conclusion}.

\section{Overview}
\label{sec-overview}
In this section, we provide an overview of \name, and discuss several critical design challenges.

\name is a failure reaction mechanism for SCR protocols. In contrast to traditional SCR protocols that specify a single path in the packet header, \name enables fast recovery within the network by allowing the source to embed the rerouting information within the packet header in the form of a {\bf forwarding subgraph (FS)}. The FS specifies a set of paths that intermediate routers can use to reroute packets in case of failures. The source, if it desires, can designate one of these paths as the {\bf primary path} to be used in the absence of failure; the rest of the paths are then treated as {\bf \deto paths} that can be used if the primary path is not available. In order to avoid forwarding loops, \name requires that the FS be a directed acyclic graph (DAG).

Performing forwarding in this way has two main benefits.  First, since the source specifies the FS, it has full control of not only the primary path, but also how the network forwards the packet when the primary path is not available.  Second, since \deto path information is embedded directly in the packet header, the network can react immediately without requiring involvement of the source, which reduces the reaction time in presence of link failures. In addition to these two benefits, the task of a router becomes simpler: a router requires only local knowledge of its neighbors, rather than needing an \deto path for every destination (which may require information such as the multi-homing locations of each host). In summary, \name achieves key benefits of SCR architectures (flexibility in route selection and scalability of network routing state) while simultaneously attaining failure reaction performance that is comparable to that of NCR architectures with backup paths.

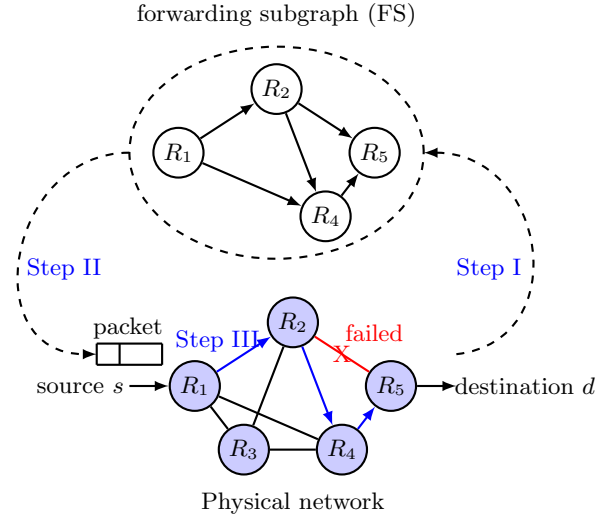
\begin{figure}[t]
		\begin{tikzpicture}[xscale=0.43, yscale=0.625]

     		\tikzstyle{sd}=[circle,draw,fill=blue!20,inner sep=2pt];
     		\tikzstyle{pnode}=[circle,draw,fill=white,inner sep=2pt];
     		\tikzstyle{snode}=[circle,draw,fill=violet,inner sep=2pt];
     		\tikzstyle{vnode}=[circle,draw,fill=red,inner sep=2pt];

      		\begin{scope}[xshift=4.25cm,yscale=0.45,>=latex]
				\node[thick] (s) at (-3.5, -0.1) {source $s$};
				\node[thick] (r1) at (0, 0) [sd] {$R_1$};
				\node[thick] (r2) at (3, 3) [sd] {$R_2$};
				\node[thick] (r3) at (1.5, -3) [sd] {$R_3$};
				\node[thick] (r4) at (4.5, -3) [sd] {$R_4$};
				\node[thick] (r5) at (6, 0) [sd] {$R_5$};
				\node[thick] (d) at (10.1, 0) {destination $d$};
		       	\node[thick] (label) at (3, -5.5) {Physical network};				

				\node[thick, red] (xx) at (4.5, 1.5) {X};
				\node[thick, red] (xx) at (5.5, 2.5) {failed};
				\draw[thick] (-3, 2) -- (-1, 2);
				\draw[thick] (-3, 1) -- (-1, 1);
				\draw[thick] (-3, 2) -- (-3, 1);
				\draw[thick] (-1, 2) -- (-1, 1);
				\draw[thick] (-2.3, 2) -- (-2.3, 1);				
				\draw[thick, ->] (-2, 0) -- (r1);
				\draw[thick, ->, blue] (r1) -- (r2);
				\draw[thick] (r1) -- (r3);
				\draw[thick] (r1) -- (r4);
				\draw[thick] (r2) -- (r3);
				\draw[thick, ->, blue] (r2) -- (r4);
				\draw[thick, red] (r2) -- (r5);
				\draw[thick] (r3) -- (r4);
				\draw[thick, ->, blue] (r4) -- (r5);
				\draw[thick, ->] (r5) -- (8, 0);
				\draw[thick, dashed, ->] (8, 1.5) to [out = 10, in = 0] (7, 11);
				\node[blue, thick] (x1) at (9, 5.5) {Step I};
				\node[thick] (x1) at (-2.05, 2.75) {packet};
				\node[thick, blue] (xx) at (0.7, 2.1) {Step III};

				\node[thick] (r11) at (-0.5, 11) [pnode] {$R_1$};
				\node[thick] (r12) at (2.5, 14) [pnode] {$R_2$};
				\node[thick] (r14) at (4, 8) [pnode] {$R_4$};
				\node[thick] (r15) at (5.5, 11) [pnode] {$R_5$};
				\draw[thick, dashed] (2.5, 11) ellipse (4.5cm and 5cm);
		       	\node[thick] (label) at (2.5, 17.5) {forwarding subgraph (FS)};

				
				\draw[thick, ->] (r11) -- (r12);
				\draw[thick, ->] (r11) -- (r14);
				\draw[thick, ->] (r12) -- (r15);
				\draw[thick, ->] (r14) -- (r15);
				\draw[thick, ->] (r12) -- (r14);
				\draw[thick, dashed, ->] (-2, 11) to [out = 180, in = 180] (-3, 1.5);
				\node[blue, thick] (x2) at (-4, 5.5) {Step II};
      		\end{scope}
   		\end{tikzpicture}
\caption{Overview of the \name design. Step I: the source selects a forwarding subgraph (FS) based on the topology of the physical network; Step II: the source encodes and embeds the FS in the packet header to inform routers how to route around encountered failures; and Step III: routers forward the packet based on the FS contained in the packet header.}
\label{fig:toy}
\end{figure}

Fig.~\ref{fig:toy} shows an example to illustrate the design of \name.  Suppose the source $s$  wishes to send a packet to a destination $d$. The source has acquired, by some mechanism to be discussed later, a map of the network. It selects the FS as shown in Fig.~\ref{fig:toy} and designates $(s, R_1, R_2,$ $R_5, d)$ to be the primary path. Note that the FS provides each node on the primary path with sufficient alternatives so that if a link on the primary path fails, the packet can be rerouted to the destination. Next, $s$ constructs a data packet with the subgraph embedded in the packet header, and forwards it on to the first-hop $R_1$. At $R_1$, the packet is forwarded to the next-hop on the primary path ($R_2$). Now suppose that at $R_2$ the primary path's next hop $R_5$ lies across a failed link $(R_2, R_5)$; then $R_2$ forwards the packet to $R_4$, the next-hop on its \deto path in the FS, after which the packet continues to $R_5$ and finally $d$.

Realizing the high level idea of source-controlled routing along an FS, however, involves several key challenges.  We outline these challenges and our solutions here.

\paragraphb{Obtaining the map}Like other SCR architectures in which sources construct end-to-end paths, our sources require a map of the available links.  When deploying \name as an interdomain routing protocol, this immediately raises questions of scalability and policy compliance.  Is it feasible to push a map of the Internet, at some level of granularity, to every source or at least every edge router?  Is there an acceptable way to balance control of network resources between the senders and the network owners?  Fortunately, we can adopt the solutions developed by past work, in particular NIRA~\cite{nira} and pathlet routing~\cite{pathlets}, which have shown how maps of policy-compliant transit service can be constructed and disseminated in ways that can be much {\em more} scalable than traditional NCR protocols like BGP.

\paragraphb{Packet header overhead} The next challenge is to design an efficient encoding mechanism that embeds the FS into the packet header with minimal overhead. By using link labels with only local significance and allocating every bit carefully, we are able to achieve acceptable packet header sizes on realistic network topologies.

\paragraphb{Fast data-plane operations}Another challenge is to design an efficient data plane forwarding algorithm: the encoding and forwarding mechanisms in \name should minimize next-hop lookup time without substantially increasing header processing cost, forwarding delay and/or design complexity of modern router forwarding planes. Fortunately, forwarding along an FS requires only lookup and pointer-increment operations, as in standard SCR protocols, and can be efficiently implemented in practice.

The next section discusses our design in more detail, including our solutions to these challenges.

\section{\nametitle Design}
\label{sec-design}

In this section, we present in more detail the four main components of \name: definition and dissemination of the network map (\S\ref{sec-design-4}); selection of a forwarding subgraph (FS) at the source (\S\ref{sec-design-1}); encoding of the FS into the packet header (\S\ref{sec-design-2}); and the data plane forwarding mechanism at routers (\S\ref{sec-design-3}).

The \name approach could be applied in multiple contexts.  We describe here how the design can be applied to interdomain and intradomain routing.  The differences principally lie in map dissemination and data plane forwarding, with the core approach taking the same form in both contexts.

\subsection{Map format and dissemination}
\label{sec-design-4}

As in other SCR protocols in which the source composes end-to-end
paths~\cite{nira, pathlets}, in \name, the source must obtain a
network ``map'' (topology) from which it can construct paths. This map
is an abstract directed graph in which each directed directed link
$(u,v)$ at node node $u$ is annotated with a {\em label}.  The label
is a compact, variable-length bitstring, which the source will use
when encoding the FS (\S~\ref{sec-design-2}) to tell node $u$ that it
wants $u$ to use the link $(u, v)$. Similar to an MPLS label, the
label identifies a link only locally at $u$, not globally.  Thus, $u$ will generally announce labels
of length $\lceil \log_2 \delta(u)\rceil$ bits where $\delta(u)$ is
the degree of $u$.

What this map corresponds to in the physical network and how the map is disseminated depend on the deployment scenario.  In an intradomain environment, the map would correspond to the physical topology of routers and links and could be distributed via a protocol like OSPF or through a centralized coordinator as in~\cite{fcp}.

In an interdomain environment, we have to deal with the significant challenges of scalability and network owners' transit policies.  In order to overcome these challenges, we build on solutions developed in past work and briefly describe them here for completeness.

\paragraphb{Basic approach}
Both NIRA~\cite{nira} and pathlet routing~\cite{pathlets} provide sources with a policy-compliant map of the Internet, roughly at the autonomous system (AS) level.  NIRA's map assumes common customer-provider-peer relationships between ASes and allows a subset of {\em valley-free} routes: that is, packets travel up a chain of providers, potentially across a peering link, and down a chain of customers to the destination. Pathlet routing represents this map explicitly as an arbitrary virtual topology, whose edges (pathlets) represent policy-compliant transit service.

\paragraphb{Scalability}
NIRA, while dependent on the existence of a typical AS business hierarchy, offers the opportunity of vastly {\em improving} BGP's control plane scalability.  Rather than learning an Internet-wide topology, each node learns its ``up-graph'' of routes through providers, stopping at the ``core'' of the Internet.  The up-graph requires fewer than 20 entries for 90\% of domains~\cite{nira}, many orders of magnitude less than the roughly 300,000 prefixes that BGP propagates today.  Each destination stores its up-graph in a global DNS-like database; to route to a destination, a source queries the database and combines its own up-graph with the destination's up-graph.  Though the resulting map is a small fraction of the Internet, it includes all policy-compliant (valley-free) routes.  Pathlet routing could use a NIRA-style approach for disseminating the pathlet topology, or it can be disseminated via a BGP-like mechanism with slightly more messaging and control state ($\leq 1.7\times$) than traditional BGP.

\name can take advantage of either the pathlet or NIRA approach for interdomain map dissemination. Thus, \name does not require a source to have complete topological knowledge of the network, but rather only enough to construct a path and alternate paths to the destination.

We also note that \name, like other SCR and multipath routing architectures, can benefit from significantly reduced rate of control plane updates~\cite{drcch10} compared with basic single-path NCR architectures. This is because short-lived failures need not be disseminated through the control plane, since failure reaction will happen anyway via forwarding along alternate paths without waiting for control-plane updates.

\paragraphb{Link labels}Along with the map itself, \name requires labels on the links.  Routers (or ASes for interdomain; for convenience we'll use ``routers'' in what follows) can piggyback this information with the link advertisements~\cite{pathlets}.  To change a label, a router readvertises the link.  While readvertisements increase control traffic, we expect that changing a router's link labels will be fairly rare, for two reasons.  First, the operator could change a single label from one bit sequence to another; however, there should be little need for such changes because the labels are arbitrary identifiers with no significance.  Second, the operator may need to increase the number of links exiting the router. This {\em may} increase the label length and require readvertisements of all of the router's link labels, creating a period of inconsistency from when the router changes its label length to when sources receive the updated announcement.  However, label lengths change
only once every time the number of outgoing links doubles (or halves) in size,
which is expected to be a very rare event.

An alternate approach is to make labels self-describing: their first few bits encode the label length~\cite{pathlets}.  This avoids the need to readvertise links after a length change and the resulting inconsistency, but labels become slightly longer.  Since compactness is important for \name, we do not evaluate this approach in this paper.

\paragraphb{Map consistency} A natural question is whether all sources and the network must have an entirely consistent view of the map at all times.  Fortunately, this difficult task is unnecessary.  There are three possible types of inconsistency.

First, if a source uses a non-existent label (e.g., the link has been removed or its label changed), this is equivalent to a link failure and the packet can be re-routed along an alternate path.  To avoid even this minor disturbance, routers can insert a short delay between announcing a label deletion and its removal from forwarding tables.

Second, if a source uses a label that has changed to identify a different link, then the packet will follow an incorrect path and will be unlikely to reach its intended destination.  This is similar to inconsistency problems in basic NCR protocols.  (Unlike in basic NCR protocols, however, the packet cannot get into a loop of any significant length because one link in the DAG will be consumed at each hop.)  To avoid label-change inconsistency, routers can simply use new labels rather than reusing ones that have recently had a different meaning.

Third, a source might be unaware of some valid labels. This simply results in a slightly restricted set of options until it receives the relevant control plane advertisement, as in essentially any other distributed routing protocol.

Thus, in all cases, inconsistency issues can be mitigated.

\subsection{Selection of the forwarding subgraph}
\label{sec-design-1}
Once a source has obtained the network map, it selects a {\em forwarding subgraph} (FS) along which it desires the packet to be routed in the network. The FS is a DAG corresponding to a subset of nodes and links in the network map. The directed edges inform routers of the packet's allowed next-hops, and acyclicity ensures there are no forwarding loops.  Additionally, for each node in the FS, the source may mark one outgoing link as the preferred primary.

Sources have a great deal of flexibility in how they choose an FS. For instance, the source may select an FS that avoids any single link failure along a low-latency primary path, avoids node failures, optimizes for other metrics like bandwidth, or picks alternate paths that avoid shared risk link groups. We discuss some of these uses in \S\ref{sec-extensions}.  For concreteness, we describe here and evaluate in \S\ref{sec-eval} how the source can pick an FS that will minimize primary-path latency and provide alternate paths to avoid any single link failure.  As noted below, accommodating shared risk link groups is similar.

A source $s$, for a given destination $d$, constructs a single-failure-avoiding FS as follows. First, $s$ computes a primary path $P$ to $d$ by running a shortest path algorithm over the network map. Next, $s$ visits each link along $P$, and computes the alternate path $P_i$ it would prefer the packet to be routed along if that link were to fail. In particular, for each node $v_i$ on the primary path, we (a) remove $v_i$'s outgoing edge corresponding to its next hop along the primary path; (b) compute a shortest path from $v_i$ to $d$, not using the removed outgoing edge; and (c) restore the removed edge. In case of a node having multiple shortest paths to the destination, the source may arbitrarily select one of these shortest paths. Finally, the primary and the \deto paths are assembled into the FS. Note that the above algorithm requires $|P|$ runs of Dijkstra's algorithm. Surprisingly, it is possible to construct a primary path and all the \deto paths in a {\em single} run of a shortest-path algorithm; see~\cite{vickery}.

Beyond single-link-failure protection, a source may want to protect against failures of shared risk link groups (i.e., sets of links that are likely to have correlated failures, such as multiple logical links allocated to a single physical fiber).  Assuming it has knowledge of these groups, it can do this by removing all links in the group in substep (a) above, and restoring them all in (c).

\begin{figure}[t]
		\centering
	\subfigure[Network map]{
		\begin{tikzpicture}[xscale=0.42, yscale=0.625]

    		\tikzstyle{sd}=[circle,draw,fill=white,inner sep=1pt,minimum width=0.6cm,minimum height=0.6cm];
    		\tikzstyle{pnode}=[circle,draw,fill=blue!20,inner sep=1pt,minimum width=0.6cm,minimum height=0.6cm];
    		\tikzstyle{snode}=[circle,draw,fill=violet,inner sep=1pt,minimum width=0.6cm,minimum height=0.6cm];
    		\tikzstyle{vnode}=[circle,draw,fill=red,inner sep=1pt,minimum width=0.6cm,minimum height=0.6cm];

     		\begin{scope}[xshift=4.25cm,yscale=0.45,>=latex]
				\node[thick] (s1) at (-0.75, 0) {$s$};
				\node[thick] (s) at (1.1, 0) [pnode] {$R_1$};
      			\node[thick] (p1) at (3.3, 2.5) [pnode] {$R_2$};
      			\node[thick] (p2) at (3.3, -2.5) [pnode] {$R_3$};
      			\node[thick] (d) at (5.5, 0) [pnode] {$R_4$};
      			\node[thick] (d1) at (7.5, 0) {$d$};

				\draw[thick, ->] (s1) -- (s);
				\draw[thick] (s) -- (p1);
				\draw[thick] (s) -- (p2);
				\draw[thick] (p1) -- (d);
				\draw[thick] (p2) -- (d);
				\draw[thick, ->] (d) -- (d1); 
     		\end{scope}
  		\end{tikzpicture}}
	\subfigure[Forwarding subgraph]{
		\begin{tikzpicture}[xscale=0.42, yscale=0.625]

    		\tikzstyle{sd}=[circle,draw,fill=white,inner sep=1pt,minimum width=0.6cm,minimum height=0.6cm];
    		\tikzstyle{pnode}=[circle,draw,fill=blue!20,inner sep=1pt,minimum width=0.6cm,minimum height=0.6cm];
    		\tikzstyle{snode}=[circle,draw,fill=violet,inner sep=1pt,minimum width=0.6cm,minimum height=0.6cm];
    		\tikzstyle{vnode}=[circle,draw,fill=red,inner sep=1pt,minimum width=0.6cm,minimum height=0.6cm];

     		\begin{scope}[xshift=4.25cm,yscale=0.45,>=latex]
				\node[thick] (s1) at (-0.75, 0) {$s$};
				\node[thick] (s) at (1.1, 0) [pnode] {$R_1$};
				\node[thick] (r1also) at (1.1, -2.5) [pnode] {$R_1'$};
      			\node[thick] (p1) at (3.3, 2.5) [pnode] {$R_2$};
      			\node[thick] (p2) at (3.3, -2.5) [pnode] {$R_3$};
      			\node[thick] (d) at (5.5, 0) [pnode] {$R_4$};
      			\node[thick] (d1) at (7.5, 0) {$d$};

				\draw[thick, ->] (s1) -- (s);
				\draw[thick, ->] (s) -- (p1);
				\draw[thick, ->] (s) -- (p2);
				\draw[thick, ->] (p1) -- (d);
				\draw[thick, ->] (p2) -- (d);
				\draw[thick, ->] (d) -- (d1);
				\draw[thick, ->] (p2) -- (r1also);
				\draw[thick, ->] (r1also) -- (p1);
     		\end{scope}
  		\end{tikzpicture}}
\caption{An FS may have multiple representations of a network map
  node, to allow ``backtracking'' without introducing cycles in the
  FS.}
\label{fig:toy1}
\end{figure}
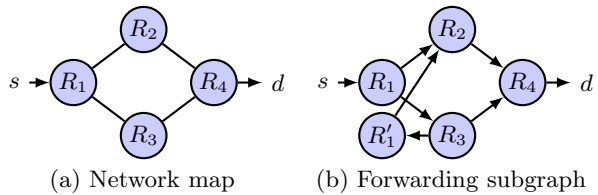

Note that there is a subtlety in how the the primary and the \deto paths are ``assembled'' into the FS: if we simply take the union of all these links and edges, we might create a loop, violating the acyclicity requirement. Consider the network map in Fig.~\ref{fig:toy1}(a). Assume that $s$ desires to use $(s, R_1, R_3, R_4, d)$ as the primary path. Then to escape a failure of the link $(R_3,R_4)$, a packet located at $R_3$ must follow the path $(R_3, R_1, R_2, R_4, d)$. Taking the union of these primary and alternate paths would result in a loop $R_1 \to R_3 \to R_1$.  Due to symmetry, the problem persists if $(s, R_1, R_2, R_4, d)$ is the primary path.

In order to avoid such loops, when adding an alternate path edge $(u,v)$ to the FS, we first check to see if this would cause a loop. If so, we create a second FS representation $v'$ of the physical node $v$, and add the edge $(u,v')$. This can be seen as ``tunneling'' the packet back along an alternate path. In the example of Fig.~\ref{fig:toy1}, before adding the second alternate path, we create a new copy $R_1'$ corresponding to the node $R_1$. The alternate path then follows $(R_3,R_1',R_2,R_4,d)$, resulting in a acyclic representation of the FS as shown in Fig.~\ref{fig:toy1}(b).

\subsection{Encoding the forwarding subgraph}
\label{sec-design-2}

After choosing an FS, the source must encode the FS into a sequence of bits and place it in the packet header. \name is agnostic to the particular location this header appears in the packet (for example, it may reside in a ``shim'' header between the IP and MAC layers, in an IP option, or in a novel header format in a next-generation Internet protocol). There are two key goals in designing an encoding format: (a) minimizing the size of the resulting encoding; and (b) ensuring data plane forwarding operations are simple. We designed and evaluated several encoding formats to achieve these goals.

In this paper we present two encoding formats, called \codecB and \codecA. Each may result in a smaller encoding in certain scenarios as discussed below. But the latter resulted in smaller encoding sizes in the network topologies we evaluated using the single-failure-avoiding FS selection (\S\ref{sec-design-1}), so it is our default.

\paragraphb{\codecB format} The \codecB format encodes the FS directly, in the sense that the FS's DAG data structure in memory is essentially directly serialized into a DAG data structure in the packet header.  The header contains a sequence of node representations, each containing one or more outgoing link representations; each link representation contains its corresponding label and a pointer to another node within the header, corresponding to the node at the other end of the link.  We describe the bit-layout of this format in detail in Appendix~\ref{sec-app-directcodec}.

\paragraphb{\codecA format} One source of overhead in the \codecB format is the use of pointers within the header.  Our \codecA format avoids some of that overhead, by grouping together sequences of labels corresponding to alternate paths, without needing an explicit representation of each node along the alternate path.  The disadvantage of this grouping is that it involves duplicating link representations, similar to how a depth-first traversal of all paths in the DAG could visit links multiple times.

In fact, there exist DAGs that have exponentially large numbers of possible traversals (thus specifying exponentially large numbers of ways the packet could be forwarded through the network).  Consequently, the \codecB format can be exponentially more efficient than the approach of \codecA in the most extreme case.  In general, we expect \codecB will be more compact for situations in which the alternate paths often share nodes with one another or with the primary.  However, in this paper we focus on the particular application of choosing single-failure-avoiding FSes.  For that application, we found that the savings from avoiding pointers outweighed the duplication of link representations, so that \codecA was somewhat more compact in several realistic networks (\S\ref{sec-eval}).  We therefore choose the \codecA format as our default and describe it in more detail now.

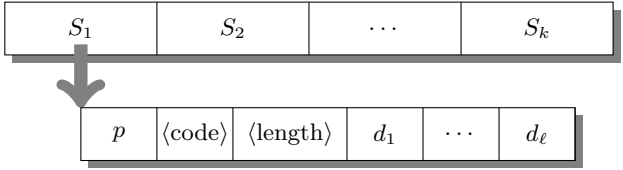
\begin{figure}[t]
\centering
\begin{tikzpicture}[yscale=0.7]
\draw[preaction={fill=black,opacity=.5, transform canvas={xshift=1mm,yshift=-1mm}}][fill=white] (0,0) rectangle (8,1); 
\draw (1, 0.5) node {$S_1$};
\draw (2, 0) -- (2, 1);
\draw (3, 0.5) node {$S_2$};
\draw (4, 0) -- (4, 1);
\draw (5, 0.5) node {$\dots$};
\draw (6, 0) -- (6, 1);
\draw (7, 0.5) node {$S_k$};
\draw[line width = 5pt, gray, ->] (1, 0.2) -- (1, -1);

\draw[preaction={fill=black,opacity=.5, transform canvas={xshift=1mm,yshift=-1mm}}][fill=white] (1,-1) rectangle (7.5,-2);
\draw (2, -1) -- (2, -2);
\draw (3, -1) -- (3, -2);
\draw (4.5, -1) -- (4.5, -2);
\draw (5.5, -1) -- (5.5, -2);
\draw (6.5, -1) -- (6.5, -2);
\draw (7.5, -1) -- (7.5, -2);
\draw (1.5, -1.5) node {$p$};
\draw (2.5, -1.5) node {$\langle {\textrm{code}} \rangle$};
\draw (3.75, -1.5) node {$\langle {\textrm{length}} \rangle$};
\draw (5, -1.5) node {$d_{1}$};
\draw (6, -1.5) node {$\dots$};
\draw (7, -1.5) node {$d_{\ell}$};
\end{tikzpicture}
\caption{\codecA encoding format layout. $S_i$ is the segment corresponding to node $v_i$ on the primary path. It encodes the node's primary next hop $p$ and \deto path $(d_{1}, d_{2}, \dots, d_{\ell})$. $\langle {\textrm{length}} \rangle$ specifies the bit-length of the \deto path, and $\langle {\textrm{code}} \rangle$ specifies the bit-length of the $\langle {\textrm{length}} \rangle$ field.
 }
\label{fig-phfn}
\end{figure}

In the \codecA format, the FS is represented as a sequence of {\bf segments}, one for each router on the primary path. For instance, in Fig.~\ref{fig-phfn}, the primary path consists of $k$ hops and $S_1, S_2, \ldots, S_k$ are the segments corresponding to those $k$ hops.  The segment corresponding to a router $v$ on the primary path contains three pieces of information (see Fig.~\ref{fig-phfn}): (a) $v$'s next-hop on the primary path; (b) the bit-length of the encoding of $v$'s \deto path; and (c) $v$'s \deto path, as a sequence of next-hop labels.  By ``$v$'s \deto path'' we mean the \deto path beginning at $v$ that avoids the primary next-hop from $v$.  (We assume here that the FS has the format of one \deto path for each link on the primary.\footnote{While the \codecA format could be generalized to have multiple alternates at a router, or segments within segments to provide alternates for routers along an alternate path, we do not explore that generalization here; in any case, such applications can use the \codecB format.})

For (a), we need to include the router's label (\S\ref{sec-design-4}) for the given outgoing edge, and similarly for (c) we include a sequence of labels. Recall that these labels are only locally unique to each node, which is critical to achieving a compact encoding, because the average number of neighbors of a router in a real-world network is typically vastly smaller than the total number of routers in the network~\cite{routeview, socialnd1}. By exploiting the structure of the real-world graphs, we are able to reduce the size of the encoding significantly compared with globally-unique labels.

\newcommand{\altlen}{$\langle {\textrm{length}} \rangle$\xspace}

For (b), we use the two fields: $\langle {\textrm{code}} \rangle$ and \altlen.  Here, \altlen specifies the total bit-lengths of all the labels $d_1, \ldots, d_\ell$ of the \deto path.  Based on our evaluation, \deto paths are shorter than 32 bits in most cases and always shorter than 128 bits; in cases a node has no \deto path, the \deto path bit-length is 0. Thus, for greater compactness, we make the bit-length of the \altlen field be variable and store it in the $\langle {\textrm{code}} \rangle$ field using a prefix-free code, with the $\langle {\textrm{code}} \rangle$ bit sequences 0, 10, and 110 mapping to values of 5, 7, and 0, respectively.

The header contains two additional pieces of information. First, the \name header begins with a two byte field, specifying its {\em header length}. Second, a one-bit field {\textsc{on-\deto?}} specifies whether the packet is traversing along the primary path or \detosingular path, and is initially false. We discuss next how routers use this information to forward packets.

\subsection{Forwarding}
\label{sec-design-3}
We now describe the forwarding mechanism used by \name routers for the \codecA format. The input to this mechanism is the \name header described in \S\ref{sec-design-2}, and the output is the interface out which the packet will be forwarded. 

Upon receiving a packet, the router first checks the value of the \name header length.  If this is $0$, this router is the destination for the packet. If not, the router checks the {\textsc{on-\deto?}} bit to see whether it is on the primary path or on \detosingular path. We describe the forwarding operations for the two cases separately.

\paragraphb{Router on the primary path} The router reads the first segment in the header, which corresponds to itself, and inspects the primary next-hop label $p$. If the corresponding link available, the router deletes this first segment corresponding to itself. It also updates the header length by subtracting the length of its segment. The packet is then forwarded to the next-hop on the primary path with the new header.

If the primary next-hop link is not available, and the \deto path length is 0, the packet is dropped. Otherwise, the router reads its next-hop label $d_1$ on the \deto path. If the link corresponding to $d_1$ is not available, the packet is dropped.\footnote{Or any other failure reaction mechanism can be applied.} If the link is available, the router removes {\em all} segments in the header, replacing them by its remaining \deto path labels $(d_2, \ldots, d_\ell)$.  It also updates the header length appropriately and sets the {\textsc{on-\deto?}} bit. The packet is then forwarded to the next-hop via label $d_1$.

\paragraphb{Router on \detosingular path} The router reads its next-hop label. If the corresponding link is not available, the packet is dropped (or, as earlier, some other failure reaction mechanism is employed). If the link is available, the router deletes its label from the header, updates the header length, and forwards the packet to the next-hop.

\paragraphb{Simplifying forwarding operations} The above description involved removing a prefix of the header, and in the case of moving to an alternate path, a suffix as well.  In some data plane implementations, these operations may be costly.  In this case, we can simply add {\em start} and {\em end} pointers at the front of the header, indicating the extent of the remaining header.  In an extra 3 bytes, we can fit two pointers that can point to individual bits in a 512-byte header (which is far larger than we need).

\paragraphb{Interdomain vs. intradomain issues} In an intradomain deployment, we may assume that each router runs \name and forwards packets as described above. However, in an interdomain deployment the forwarding subgraph roughly represents AS-level paths (as discussed in more detail in~\ref{sec-design-4}). When the packet is forwarded though an intermediate domain, that domain must forward the packet on to the next AS-level hop. Network operators may independently choose from a variety of ways to do this, for example by tunneling the packet with MPLS, or perhaps running \name internally as well as interdomain.

\section{Evaluation}
\label{sec-eval}
\name advocates the idea of embedding a forwarding subgraph (FS) in the
packet header, giving routers multiple forwarding options in order to
provide the source with some property that it desires. While \name can
support flexible FS selections that provide different guarantees, for concreteness, this
section evaluates the FS selection exemplified in \S\ref{sec-design-1}, which targets fast reaction in the
presence of single-link failures.  The source constructs a DAG
comprised of the shortest primary path, and the shortest \deto path
for each node on the primary path in case that node's outgoing link
along the primary path fails. In terms of performance, three metrics
are important: (a) encoding size, (b) failure reaction effectiveness,
and (c) router complexity and packet forwarding rates. We present
results for (a) and (b) in this section and discuss (c) in
\S\ref{sec:conclusion}.

\paragraphb{Topologies} We use three network topologies in our
evaluation: the latency-annotated topology from Sprint ISP 1239
\cite{rocketfuel}, with $315$ nodes and $972$ links; an AS-level map
of the Internet \cite{routeview}, with 33,508 nodes and 75,001
links; and the largest component, with 190,914 nodes and 607,610
links, of a router-level map of the Internet \cite{caida}. The latter
two topologies lack latency information; we take all links to have
equal length.  While using \name directly on a router-level map of the
Internet is not a likely deployment scenario (due to privacy and
scaling issues, ASes do not propagate internal topologies globally in
today's Internet), we consider this extreme design point to
investigate scaling issues of our design.

\begin{figure*}[ht]
  \centering
  \subfigure[Sprint Topology]{
    \includegraphics[width = 1.9in, height = 1.4in]{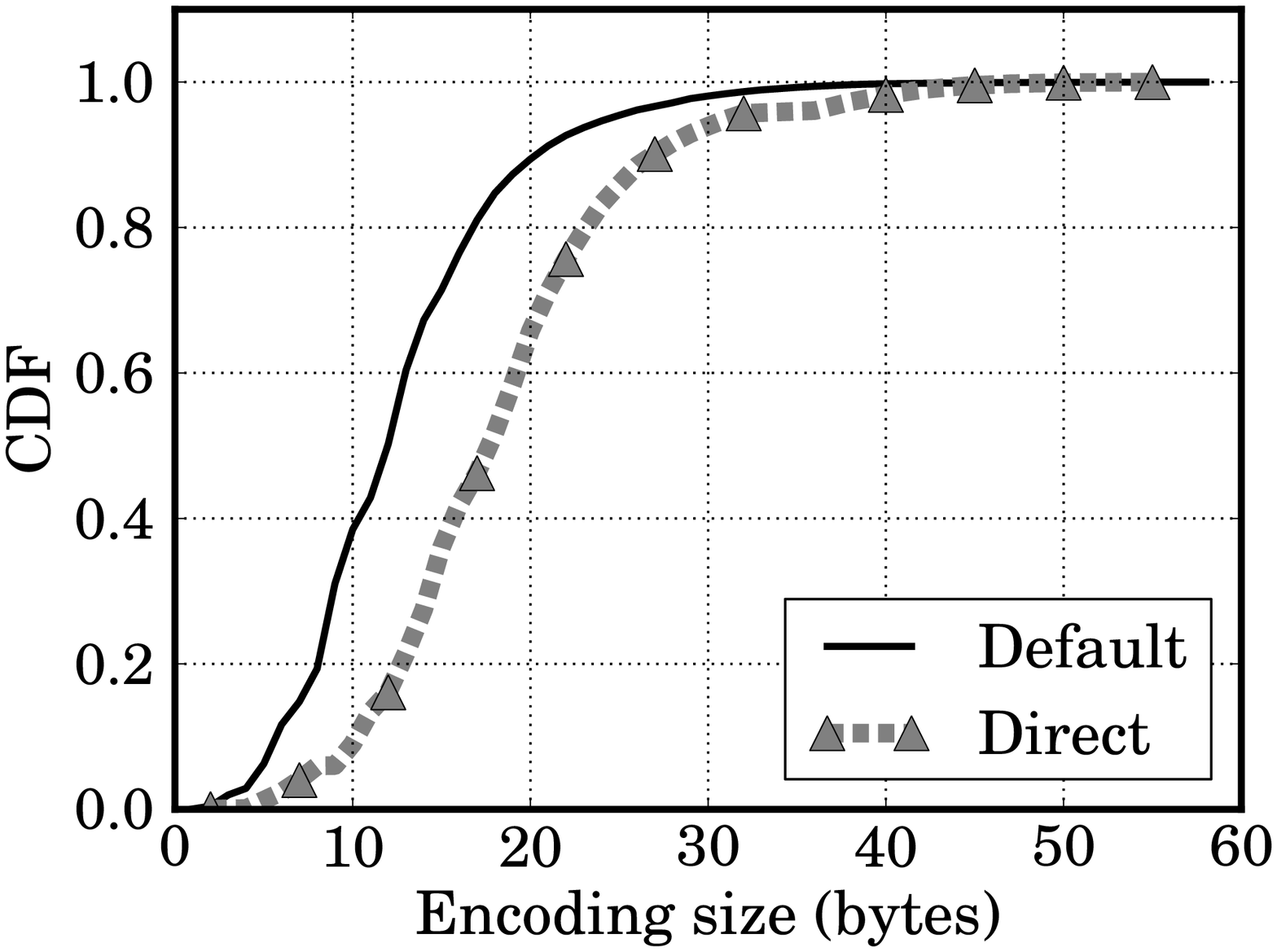}
    \label{fig:hdrsizeCDF_rocketfuel_1239}
  }
  \subfigure[AS-level Topology]{
    \includegraphics[width = 1.9in, height = 1.4in]{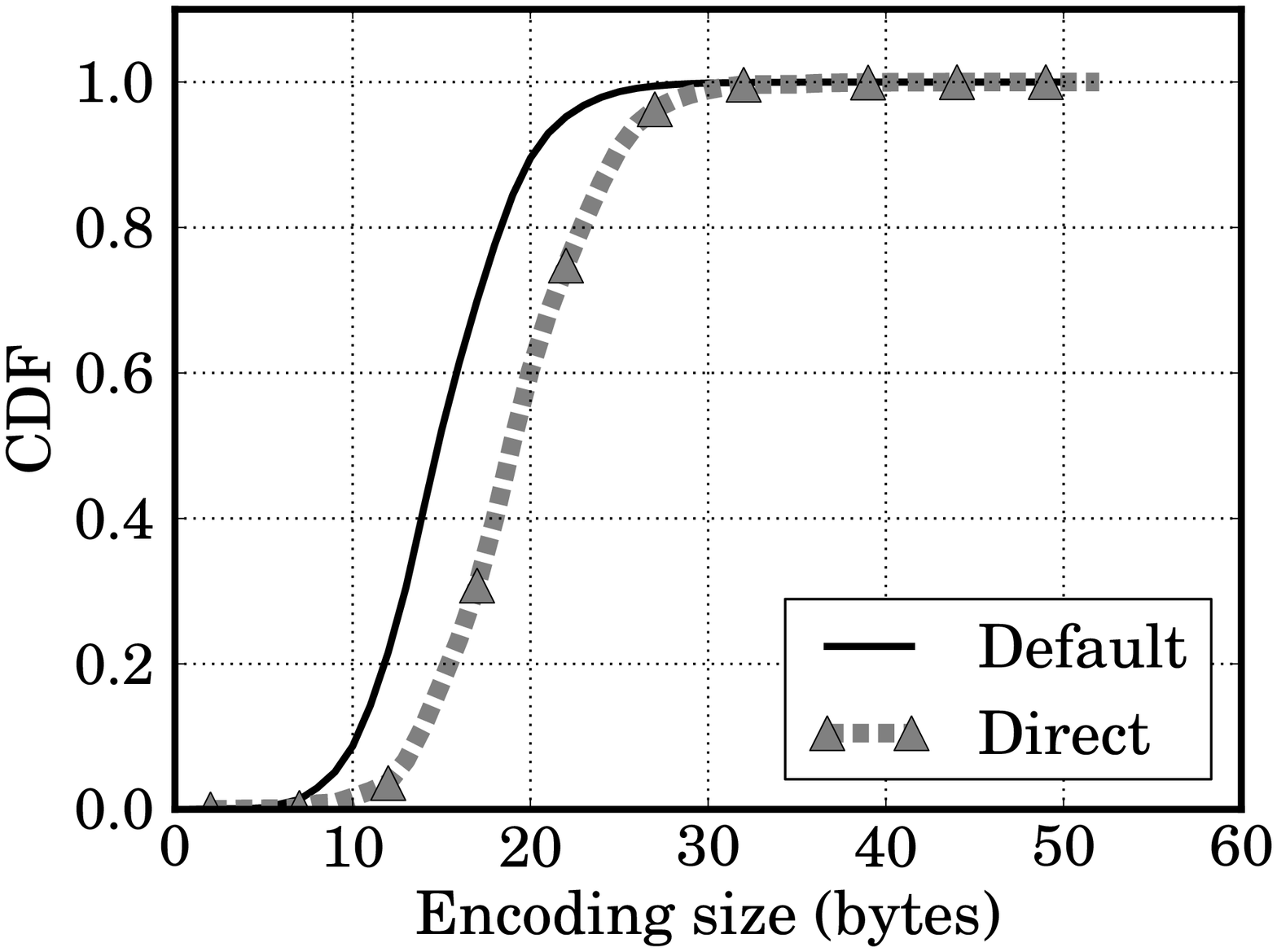}
    \label{fig:hdrsizeCDF_caidaas}
  }
  \subfigure[Router-level Topology]{
    \includegraphics[width = 1.9in, height = 1.4in]{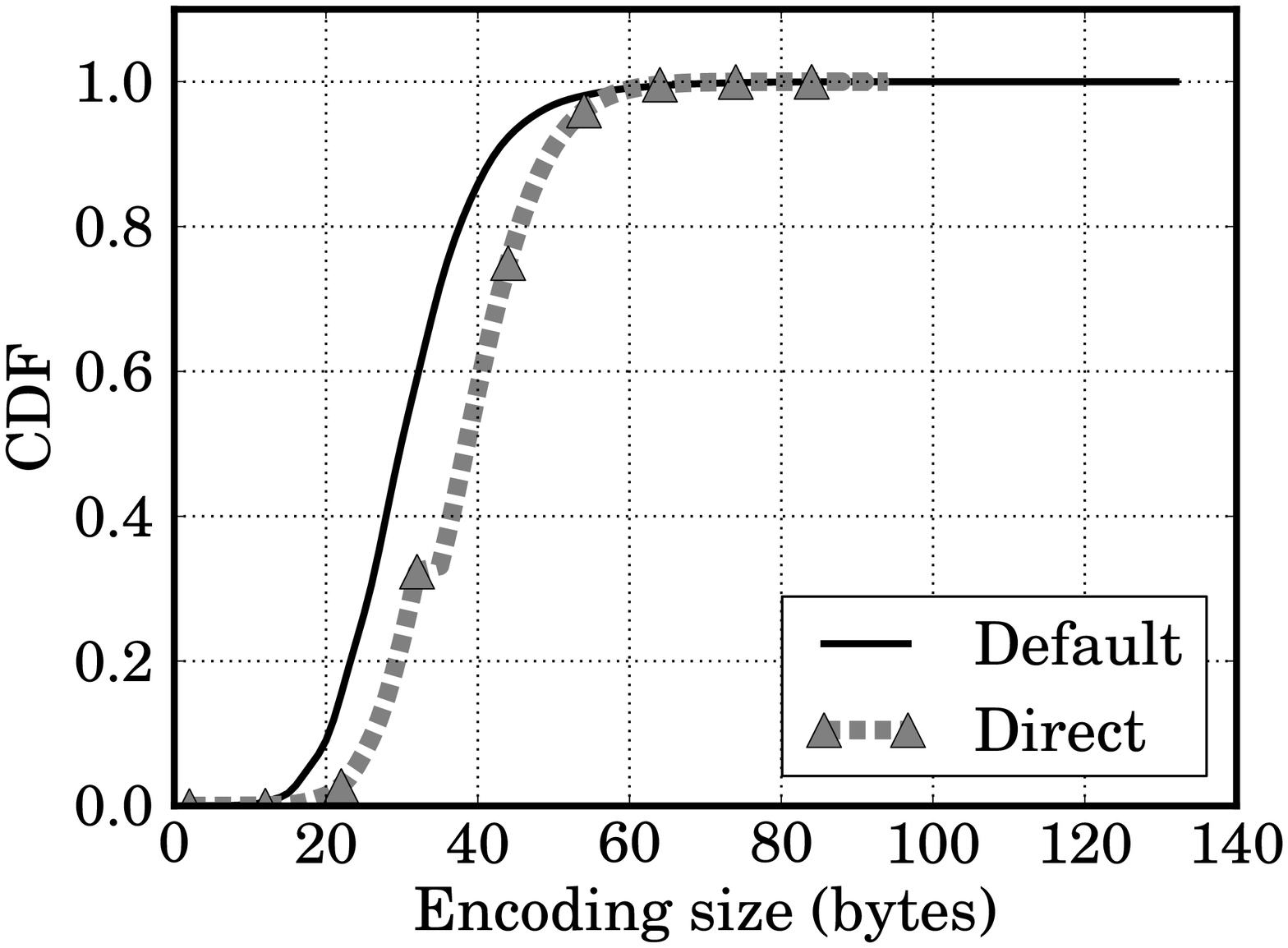}
    \label{fig:hdrsizeCDF_caida}
  }
  \caption{CDF of \name encoding size in bytes for the Direct and Default encoding formats, for handling single-link failures. 
}
  \label{fig:hs}
\end{figure*}

\begin{figure*}[ht]
  \centering
  \subfigure[Sprint Topology]{
    \includegraphics[width = 1.9in, height = 1.4in]{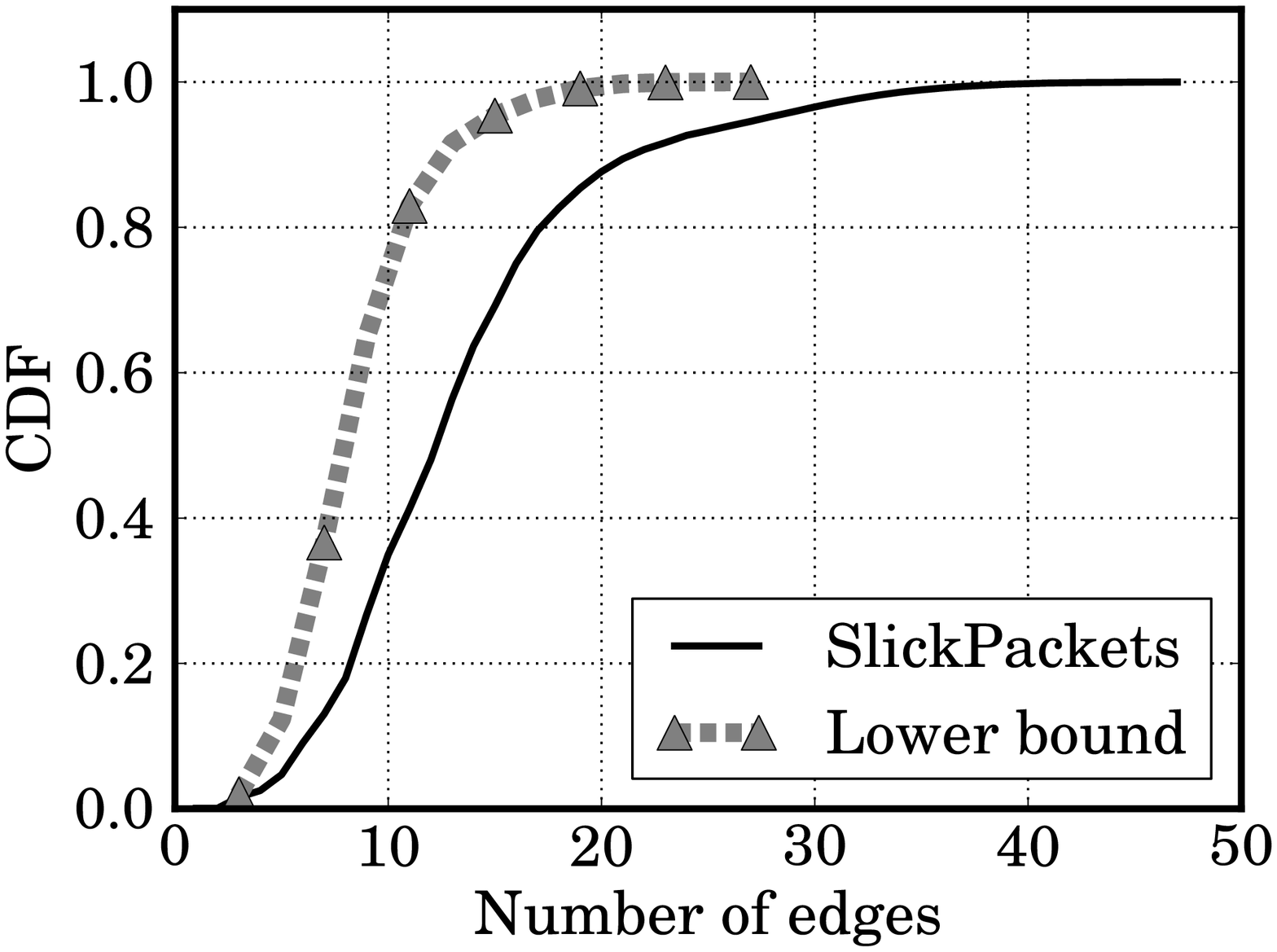}
    \label{fig:hdrsizeCDF_rocketfuel_1239}
  }
  \subfigure[AS-level Topology]{
    \includegraphics[width = 1.9in, height = 1.4in]{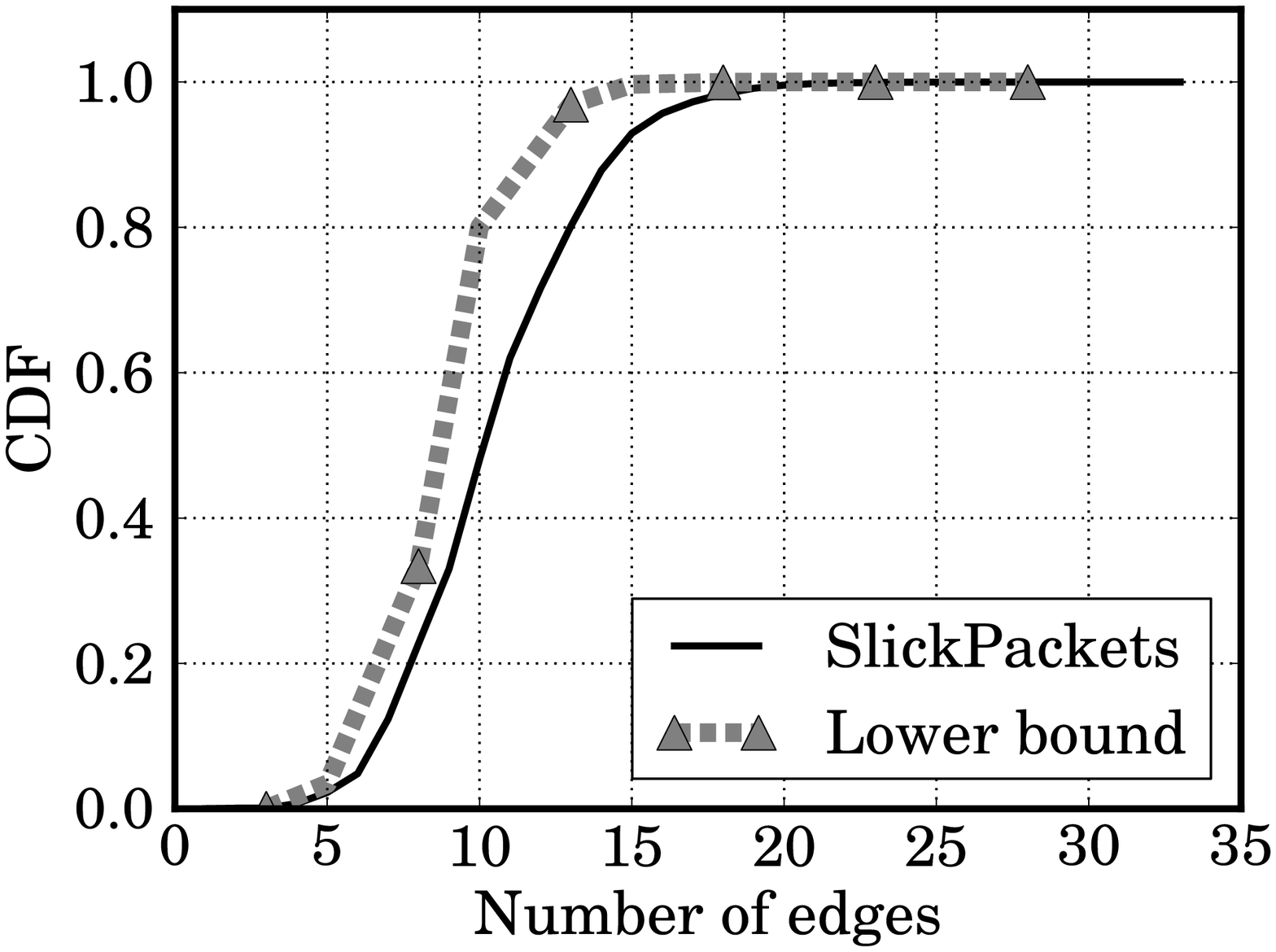}
    \label{fig:hdrsizeCDF_caidaas}
  }
  \subfigure[Router-level Topology]{
    \includegraphics[width = 1.9in, height = 1.4in]{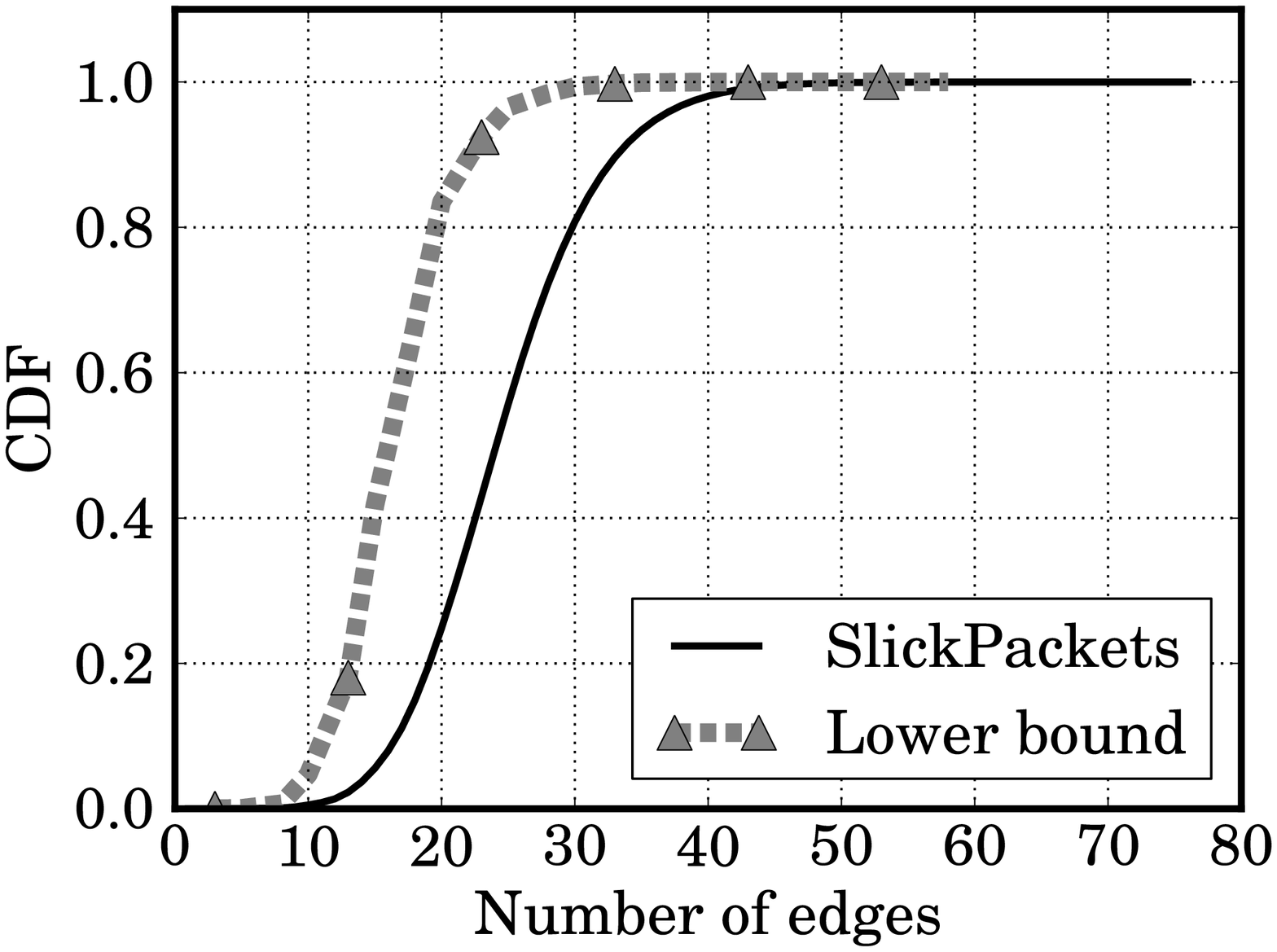}
    \label{fig:hdrsizeCDF_caida}
  }
  \caption{CDF of \name FS size and the lower bound in number of edges for
    handling single-link failures.}
  \label{fig:lab}
\end{figure*}

\subsection{Encoding size}
\label{ssec-es}
Since we encode the FS into the packet header, the
\textbf{encoding size} determines the bandwidth overhead. We evaluate
the resulting encoding sizes of the \codecB and \codecA encoding
formats presented in \S\ref{sec-design-2}, for FSes
constructed using the algorithm presented in
\S\ref{sec-design-1}.

 Furthermore, regardless of the encoding format used, the
\textbf{FS size}---the number of edges---is a factor influencing the
encoding size. We are thus also interested in comparing the sizes of FSes
constructed by the algorithm described in \S\ref{sec-design-1} to \textbf{lower bounds} on the
sizes of FSes returned by any algorithm that provides shortest path
latencies and single-link failure protection. These lower bounds
impose a fundamental limit on the encoding size; intuitively, for a given encoding format that already uses optimized label lengths, it is
hard to reduce the encoding size significantly without reducing the
FS size. We describe in Appendix~\ref{sec-app-lowerboundalgo} an
algorithm that yields a lower bound on the size of the FS for a given primary path hopcount.

\paragraphb{Methodology}We evaluate all 98,910 possible ordered
source-destination pairs of the Sprint topology. For the AS-
and router-level topologies, we randomly sample ten million unique
ordered source-destination pairs. For each pair, we
record these values: the \codecA and \codecB encoding sizes,
the size of the FS constructed using our algorithm, and the lower
bound on FS sizes.

\paragraphb{Results}Fig. \ref{fig:hs} shows the encoding size
results. We see that \codecA has
somewhat smaller size almost always; \codecB performs noticeably better only in the extreme tail of the router-level topology.  We therefore discuss \codecA in what follows.
For the intradomain Sprint topology, the maximum
encoding size is $58$ bytes. The plot has a long tail with $90\%$ and
$99\%$ of the source-destination pairs requiring less than $21$ bytes
and $34$ bytes of encoding, respectively. For the interdomain
AS-level map of the Internet, the maximum encoding size is $50$
bytes. As with the Sprint topology, the plot has a long tail, with
$90\%$ of the source-destination pairs resulting in encodings of less
than $21$ bytes; $99\%$ of the source-destination pairs result in less
than $26$ bytes.

For the extreme case of router-level topology, $90\%$ of
the source-destination pairs result in encodings of less than $43$
bytes; $99\%$ less than $60$ bytes. The remaining less than $1\%$ of
the source-destination pairs constitute the long tail, with maximum
encoding size of $132$ bytes. Although the router-level realization of
\name may be impractical, the above results demonstrate that \name can
scale on graphs as large as 200,000 nodes with moderate increase in
the packet header sizes.  If desired, this overhead may be amortized
over more data (e.g., by leveraging IPv6 jumbo frames) or using \name
only for application data that is most sensitive to failures.

Fig. \ref{fig:lab} shows the FS size (in number of links) and lower
bound. For the AS-level and router-level topologies, our FS size is
very close to the lower bound; for the Sprint topology, the difference is somewhat larger. Overall, the results suggest that, for handling
single-link failures, our simple FS selection algorithm is relatively close to optimal in terms of minimizing the number of links in the FS.

For the Sprint topology, there is also a long tail in both our FS
sizes and the lower bounds. The reason is that there are a few source-destination pairs
that have long primary paths, requiring \deto paths for a large
number of nodes, resulting in larger number of edges.

\subsection{Failure reaction effectiveness}
\label{ssec-stretcheval}
One metric to evaluate the effectiveness of a failure reaction
mechanism is the packet \textbf{stretch}, the ratio of the length of a
packet's path to the length of the shortest possible path. Previous
works calculate stretch based on packets' traversed path costs or
transit times. However, for a delay-sensitive application, we are
interested in the time a packet is \emph{live} from the application's
perspective---from the time the packet is generated by the source
application to the time it is received by the destination. Thus, we
define the stretch for a packet that does not fully traverse the
original shortest path, to be the ratio of the time the packet is
\emph{live} to the post-link-failure shortest path latency; for other
packets---those that traverse the original shortest path---the stretch
is 1. For brevity of the ensuing discussion, $l_0$ denotes the failed link
on the primary path from source $s$ to destination $d$; $r_0$ denotes
the router that is adjacent to and upstream from $l_0$ on the primary
path; and $t_0$ denotes the time of failure of $l_0$.

\paragraphb{Modeling delay at network devices}A router in the
network, upon a link failure, has to perform a number of tasks before
it has new valid default next hops for affected destinations. The four
major tasks are: (1) detecting a failed link (if the router is
adjacent to the failed link) and generating a control plane message;
(2) processing of received control packets; (3) computing the new shortest path tree (SPT);
and (4) updating the forwarding information base (FIB). We assume that
the delay in detecting a failed link is zero since irrespective of the
underlying routing architecture, all packets during this period are
lost;\footnote{Unless packets are duplicated along multiple paths---a design point that may be reasonable for certain kinds of traffic, but which we do not consider in this paper.} this does not make a difference in our performance comparison
results. We consider the three other major contributors.

Let $d_r$ be the time spent by a router in processing a control packet
(i.e., the time between the router's receipt and forwarding of the
packet). $d_r$ (along with link latencies) dictates the propagation
rate of control packets through the network. Let $d_p$ be the delay
between a router's learning of the link failure and starting a new SPT
computation; $d_c$ be the time taken to compute the new SPT; and $d_u$
be the time taken to update the FIB. Note that, upon receiving a
control packet, a router necessarily spends $D=(d_p + d_c + d_u)$ time
before having new valid default next hops for affected
destinations. The values of $d_c$ and $d_u$ depend on the router
architecture, algorithms in use, the topology, and the router's
location.  Lacking a
good model, we set these values to 0  in our simulations. However, we
use $D=d_p = 50$~ms \cite{safeguard} and $d_r = 2$~ms \cite{delay1,
  delay2} for the Sprint topology. For the AS-level and router-level
topologies, we use $D=d_r=0$.

\subsubsection{Failure reaction schemes}
\label{ssec-schemes}
The performance of source routing protocols also depends
on the \emph{control plane} mechanism: the technique used to inform
sources about the failures in the network. We describe three variants
of \name design with different control plane mechanisms. We also
describe three protocols---one from the SCR paradigm and two from the NCR
paradigm---that we compare with \name.

\paragraphb{Flooded-\name}Upon detecting the link failure, $r_0$ floods the network with a
link state advertisement (LSA). This is similar to running an SCR protocol
with an OSPF \cite{ospf} style control plane mechanism.

\paragraphb{Fast-\name}When $r_0$ receives a packet whose primary
next-hop traverses $l_0$, it informs $s$ about
the link failure by directly sending an ICMP-style notification message to
$s$. The rationale is that, to reduce control overhead, only sources
that use $l_0$ in their primary paths need to be
notified. Intuitively, this significantly reduces the control plane
packets sent into the network. 

\paragraphb{e2e-\name} The router $r_0$ piggybacks the link-failure information on the packet being forwarded on the \deto path towards $d$, which, upon receiving this information, may inform $s$ of the link failure. Thus, failure information is sent to the source in an end-to-end manner.

All \name schemes use the same FS selection algorithm
(\S\ref{sec-design-1}) and incur the delay $D$ between learning of the failure and switching to new primary paths.

\paragraphb{Vanilla source routing (VSR)} For purposes of comparison
with \name, we evaluate a simple ``vanilla'' source routing protocol.
In VSR, each source $s$ specifies a single shortest path to
its destination $d$ in the packet header. For the control plane
mechanism, we use the ``fast'' version, where $r_0$ directly
notifies $s$. After receiving the notification, $s$ incurs the delay $D$ before
computing a new shortest path. Without a valid path, packets generated during this time are queued. Packets that use $l_0$ in their paths will be
  dropped by $r_0$ after the link failure.  However, once $s$ has computed a new path, it resends the packets that would have been dropped, i.e.,
those that it sent in the time interval $[t-R,t)$ where $t$ is the time $s$ learned of the failure, and $R$ is the RTT between $s$ and
  $r_0$. Note that for some of these resent packets, there could be two
concurrent live copies: the resent copy that will be delivered along
the new path, and the original copy that will be dropped when
it reaches $r_0$.  This scheme may be difficult or undesirable to implement in practice, but as an idealized VSR, it is a useful comparison.

\paragraphb{Ideal-\sg} We simulated an idealized version of
\sg~\cite{safeguard}, a network-controlled routing protocol that
achieves fast failure reaction. \sg uses the standard OSPF as
  the control plane substrate. In \sg, $r_0$ immediately uses
  pre-computed shortest alternate paths to quickly redirect packets
  that it would otherwise forward along $l_0$. Other routers recognize
  redirected (``escort mode'') packets and forward them along their
  intended alternate paths; however, until they have updated their
  FIBs (after delay $D$ after receiving the LSA), these routers
  continue to forward ``normal mode'' packets along their sub-optimal
  paths towards $l_0$. In practice, the ``alternative path
  databases,'' which are found to be 2 to 8 times larger than a
  router's intradomain FIB \cite{safeguard}, might increase lookup latencies or
  be an impractical memory requirement. However, our ideal
  version of \sg ignores these issues.

\paragraphb{Ideal-NCR} This represents an ideal (and unachievable) NCR
scheme, in which each router learns of a link failure in exactly the
propagation delay along the shortest path from the point of failure to
the router; and the router instantly begins forwarding
packets along the shortest alternate path. Ideal-NCR is
equivalent to a special case of Ideal-\sg where all delays, except
propagation delay, are zero (i.e, $D=d_r=0$).

\subsubsection{Methodology}
\label{ssec-simulator}

We wrote a static simulator for our evaluation purposes. The simulator uses the packet stretch
computations described in Appendix~\ref{sec-app-computestretch}. Since we are evaluating
the reaction to single-link failures, we evaluate only
$(l_0, s, d)$ triples where the primary path from $s$ to $d$ uses
$l_0$, and $s$ and $d$ remain connected after the failure of
$l_0$, so that at least one \deto
path to $d$ exists for each router upstream from $l_0$. For the
Sprint topology, we evaluate all 424,569 possible
such triples. For each of the AS- and router-level
topologies, we sample 1,000 random links and use a sampling algorithm (described in Appendix~\ref{sec-app-samplingalgo})  to obtain over 750,000 and 890,000 such triples, respectively.

In our simulations, the application at the source generates packets
every 1~ms, starting at time $t=0$~ms. For the time of link failure
$t_0$, however, recall that in Ideal-\sg, Ideal-NCR, and Flooded-\name, $r_0$
floods the LSA when it detects the link failure,
not when it receives sources' packets. For these schemes, the
sooner the link fails, the sooner intermediate routers and the source
learn of the failure and use better paths. So, for a fair
comparison with non-flooding schemes, we consider two extreme points:
when $t_0$ is greater than the network diameter in
terms of \emph{link latencies} and when $t_0 = 0$. The former case
ensures that by the time $t_0$, all sources in all
evaluated $(l_0, s, d)$ triples have had packets reaching
$r_0$. For the Sprint topology, with a diameter of 139~ms, we
  use $t_0=150$. For the AS- and router-level topologies, we assume
  all links have latencies 1~ms and use $t_0=50$.

\begin{figure*}
  \centering
  \subfigure[Sprint Topology]{
    \includegraphics[height = 1.4in]{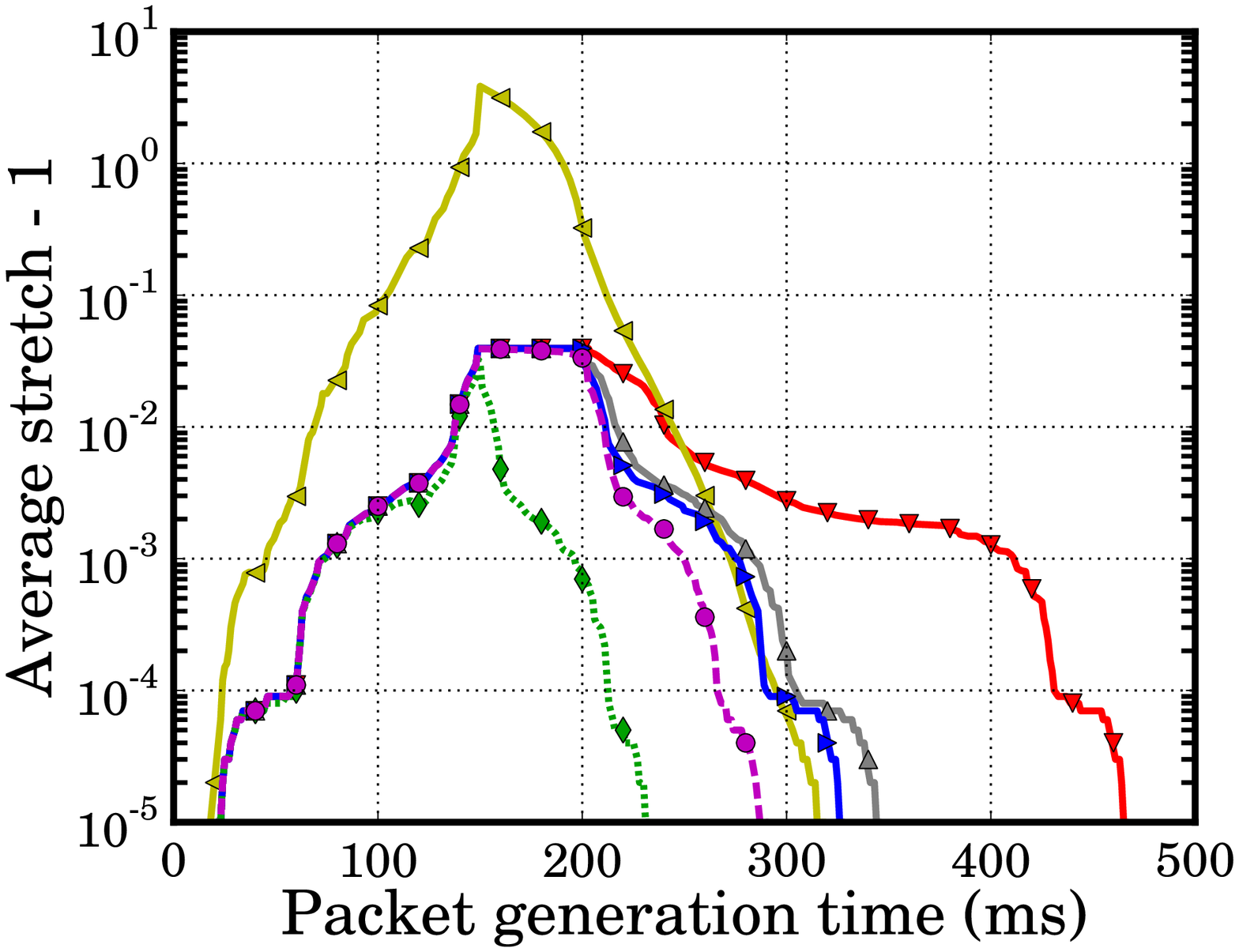}
    \label{fig:avgsvst_t0_is_diameter_rf1239}
  }
  \subfigure[AS-level Topology]{
    \includegraphics[height = 1.4in]{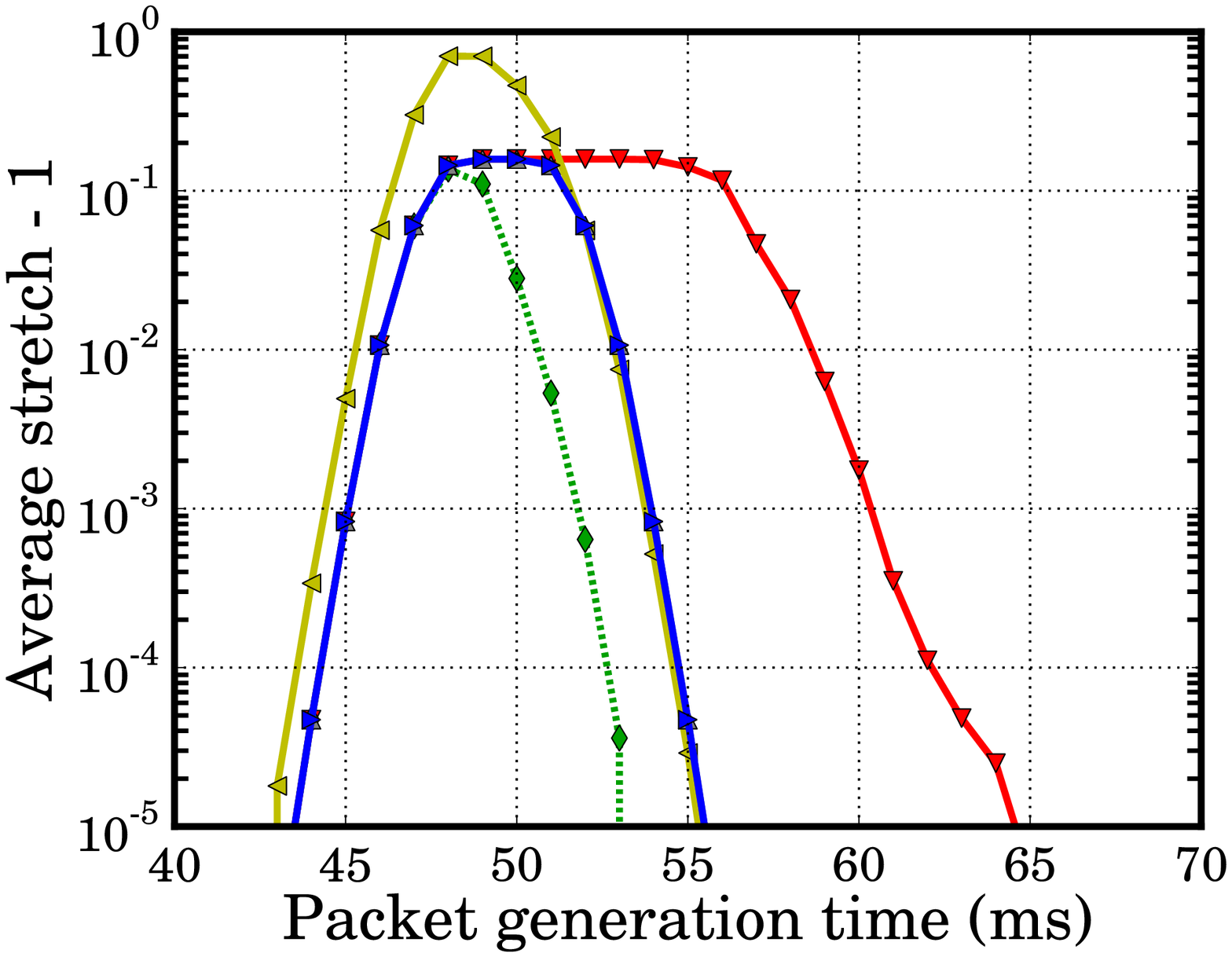}
    \label{fig:avgsvst_t0_is_diameter_aslevel}
  }
  \subfigure[Router-level Topology]{
    \includegraphics[height = 1.4in]{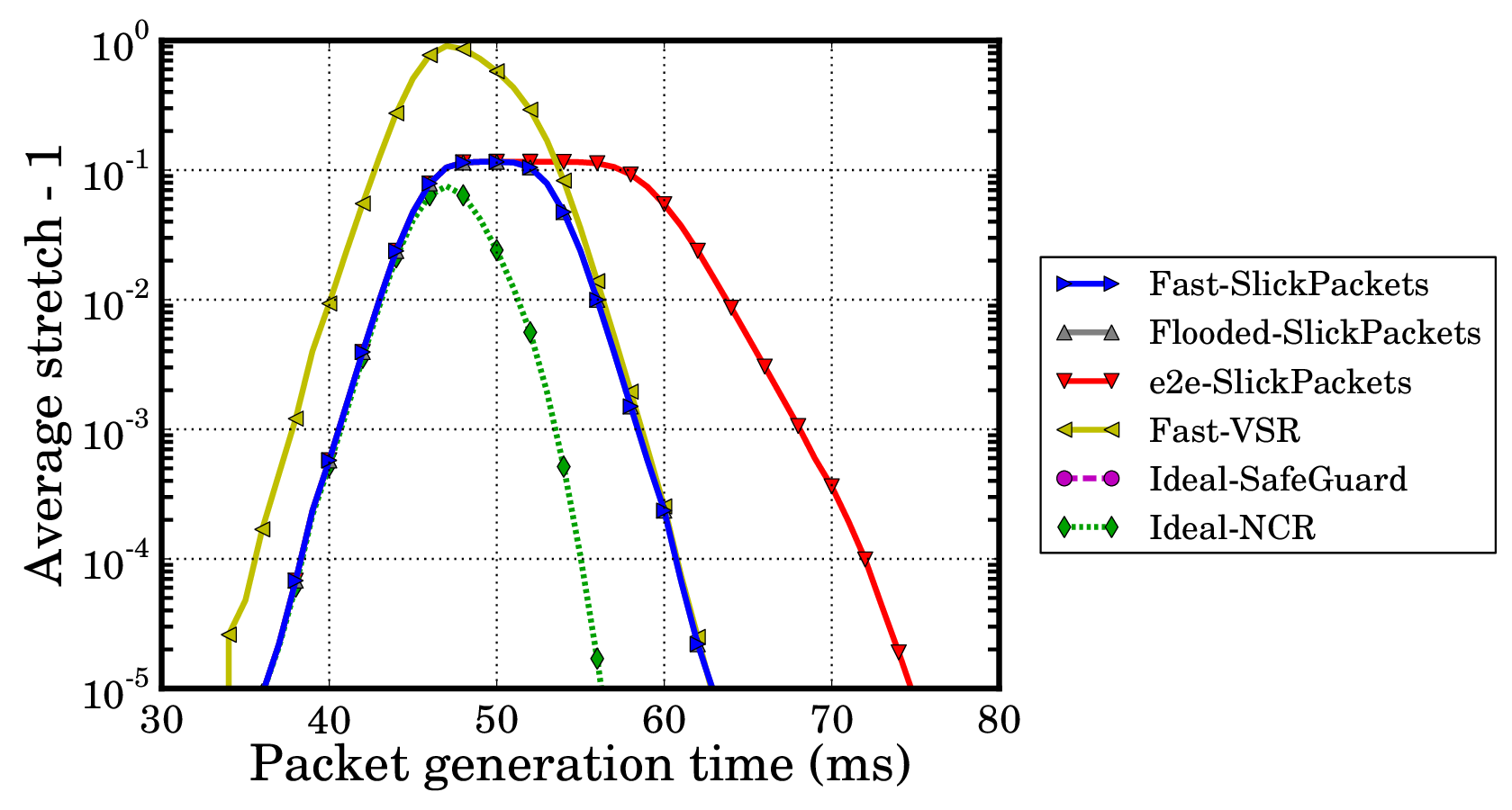}
    \label{fig:avgsvst_t0_is_diameter_routerlevel}
  }
  \caption{Average packet stretch - 1 vs. packet generation time when
    $t_0$ is greater than the network diameter. The y-axes are on log
    scales. For the Sprint topology, $t_0=150,D=50,d_r=2$. For the
    AS- and router-level topologies, $t_0=50,D=d_r=0$.}
  \label{fig:avgsvst_t0_is_diameter}
\end{figure*}
\begin{figure*}
  \centering
  \subfigure[Sprint Topology]{
    \includegraphics[height = 1.4in]{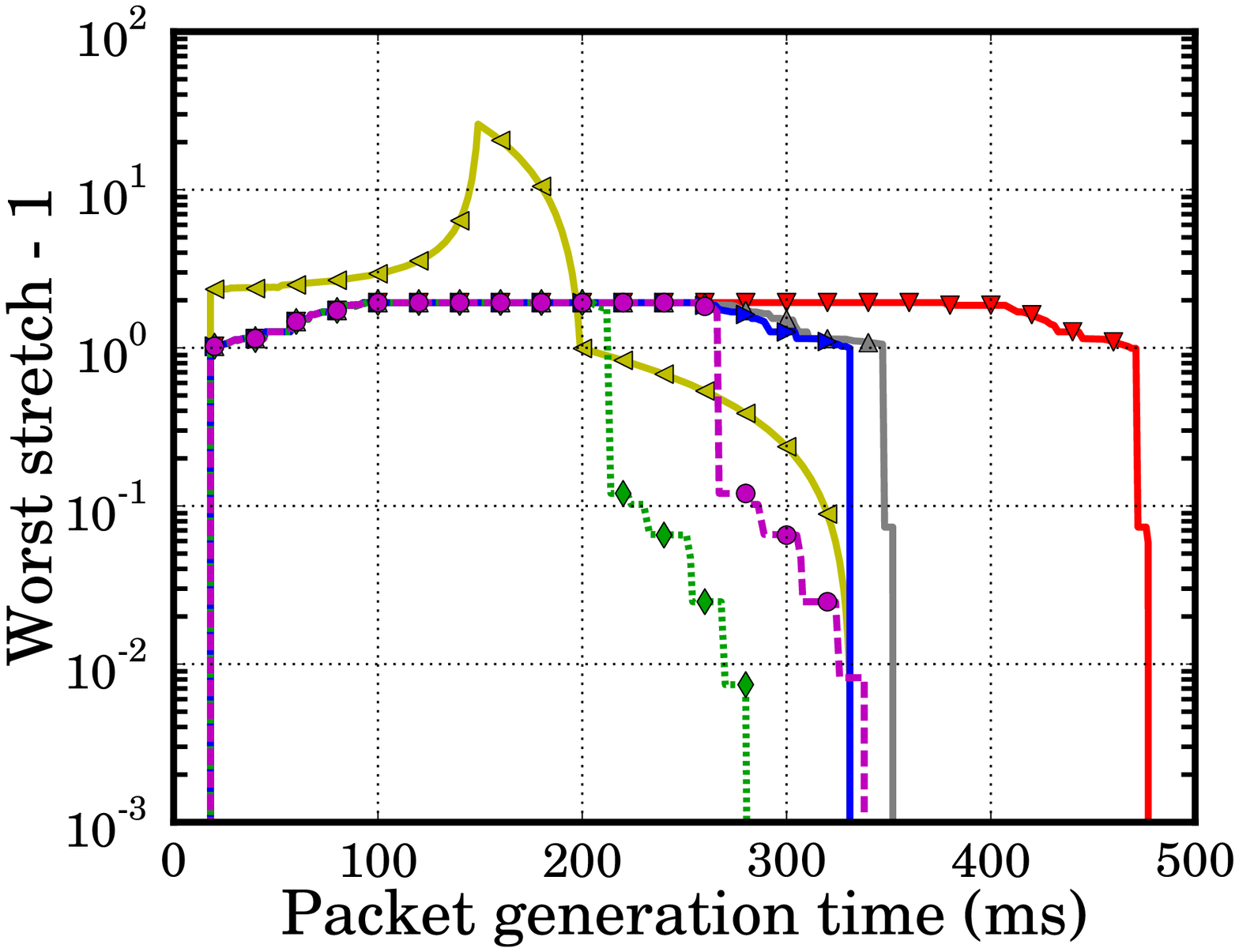}
    \label{fig:maxsvst_t0_is_diameter_rf1239}
  }
  \subfigure[AS-level Topology]{
    \includegraphics[height = 1.4in]{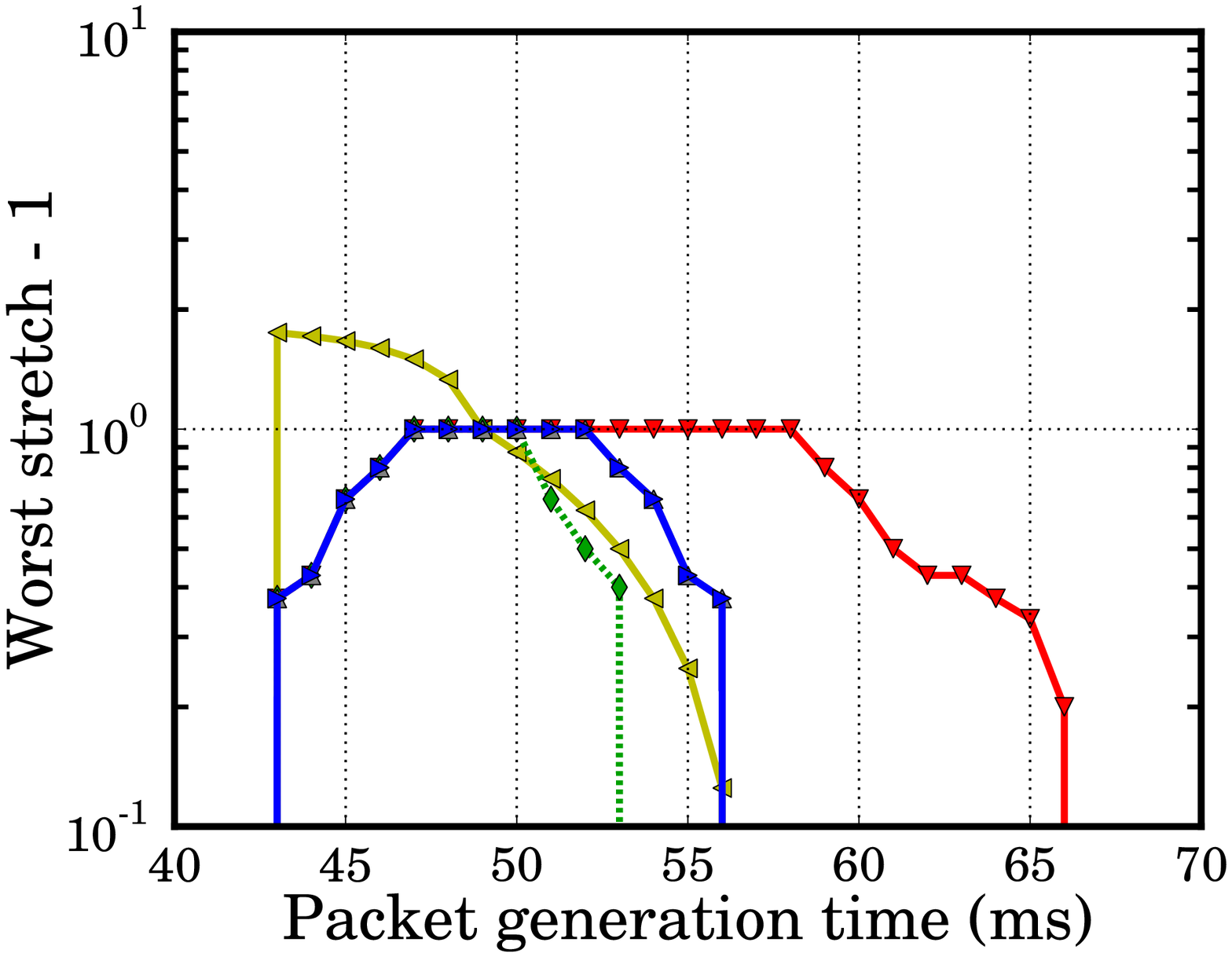}
    \label{fig:maxsvst_t0_is_diameter_aslevel}
  }
  \subfigure[Router-level Topology]{
    \includegraphics[height = 1.4in]{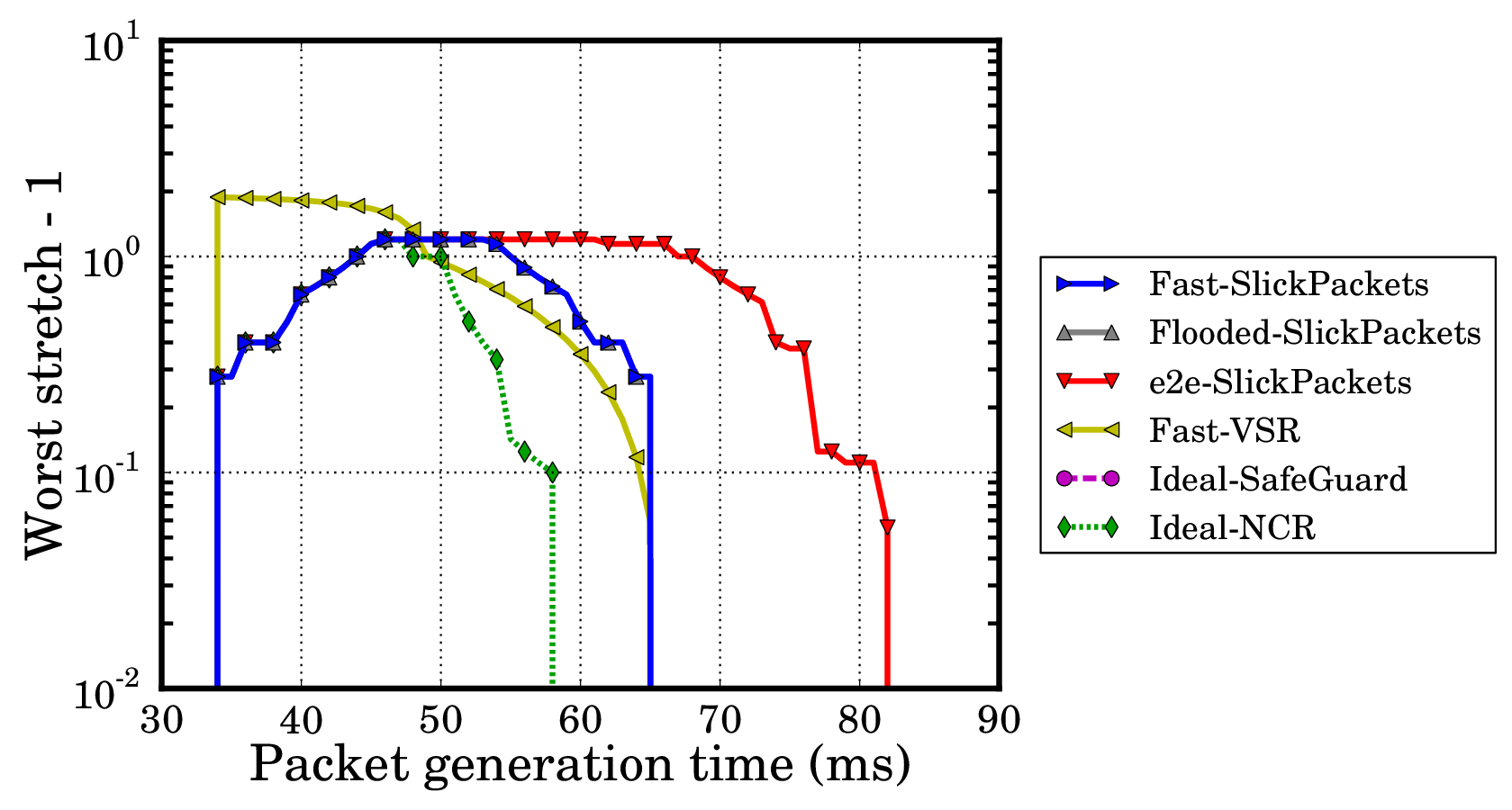}
    \label{fig:maxsvst_t0_is_diameter_routerlevel}
  }
  \caption{Worst packet stretch vs. packet generation time when
    $t_0$ is greater than the network diameter. The y-axes are on
    log scales. For the Sprint topology,
    $t_0=150,D=50,d_r=2$. For the AS- and router-level topologies,
    $t_0=50,D=d_r=0$.}
  \label{fig:worst_t0_is_diameter}
\end{figure*}
\begin{figure*}
  \centering
  \subfigure[Sprint Topology]{
    \includegraphics[height = 1.4in]{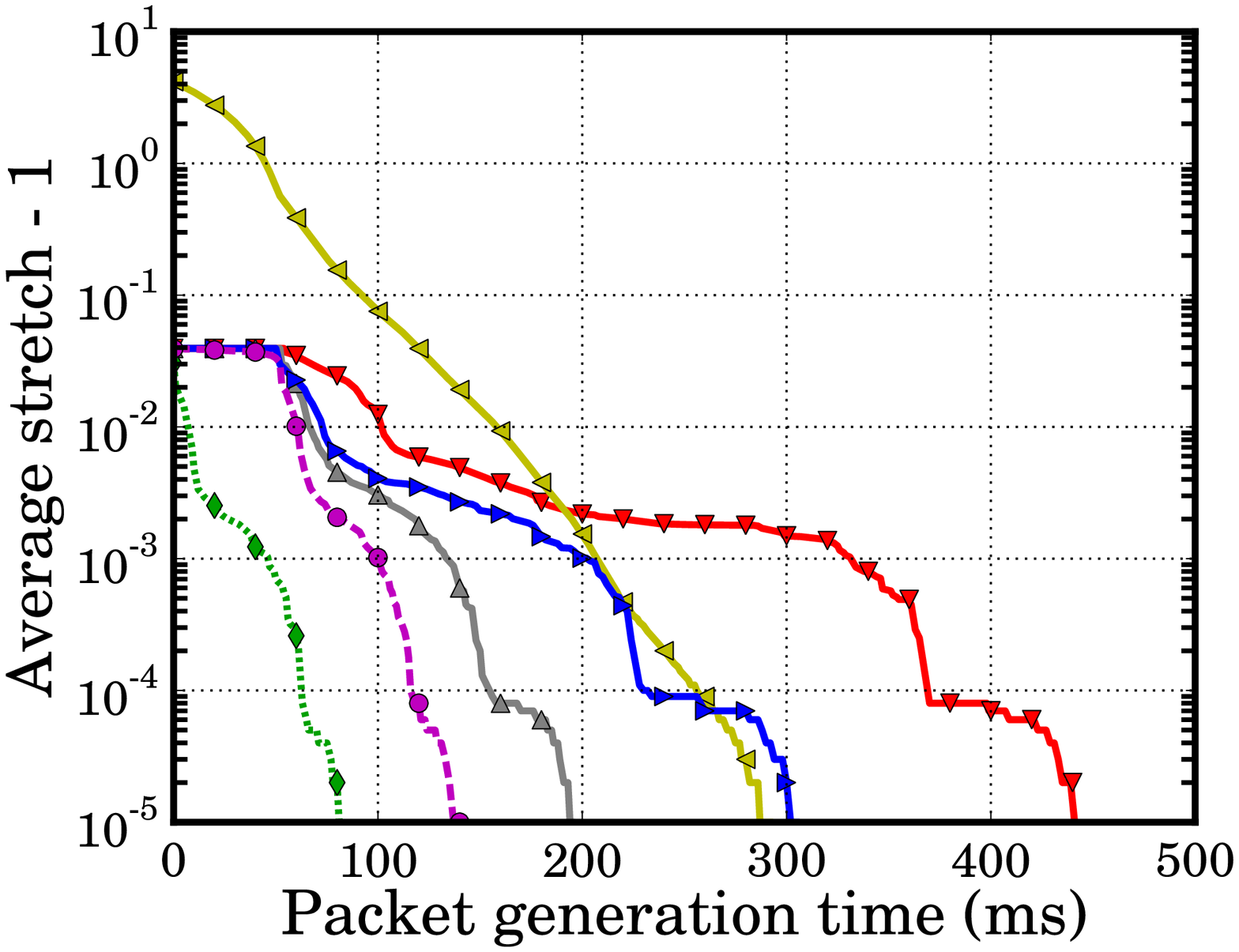}
    \label{fig:avgsvst_t0_is_0_rf1239}
  }
  \subfigure[AS-level Topology]{
    \includegraphics[height = 1.4in]{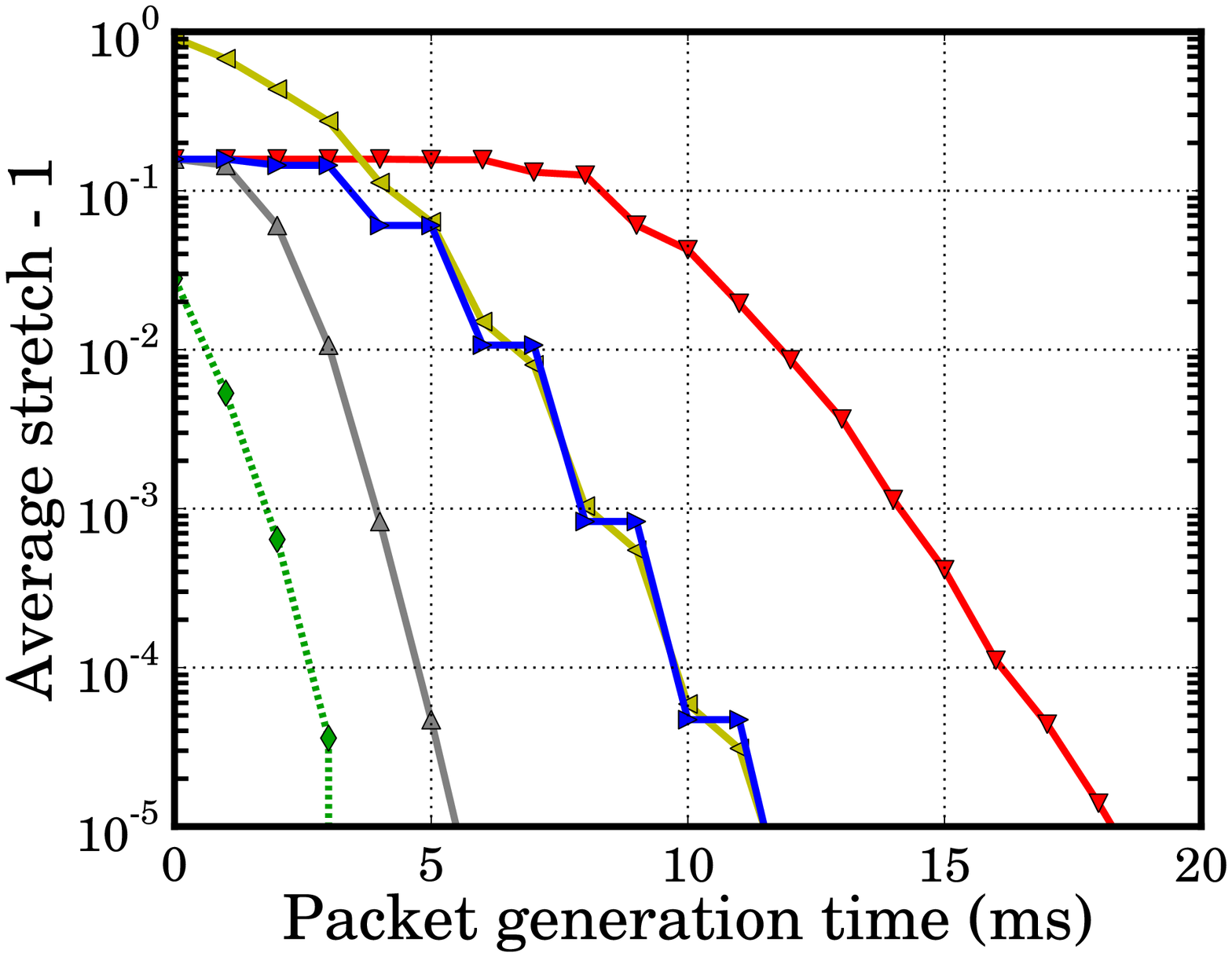}
    \label{fig:avgsvst_t0_is_0_aslevel}
  }
  \subfigure[Router-level Topology]{
    \includegraphics[height = 1.4in]{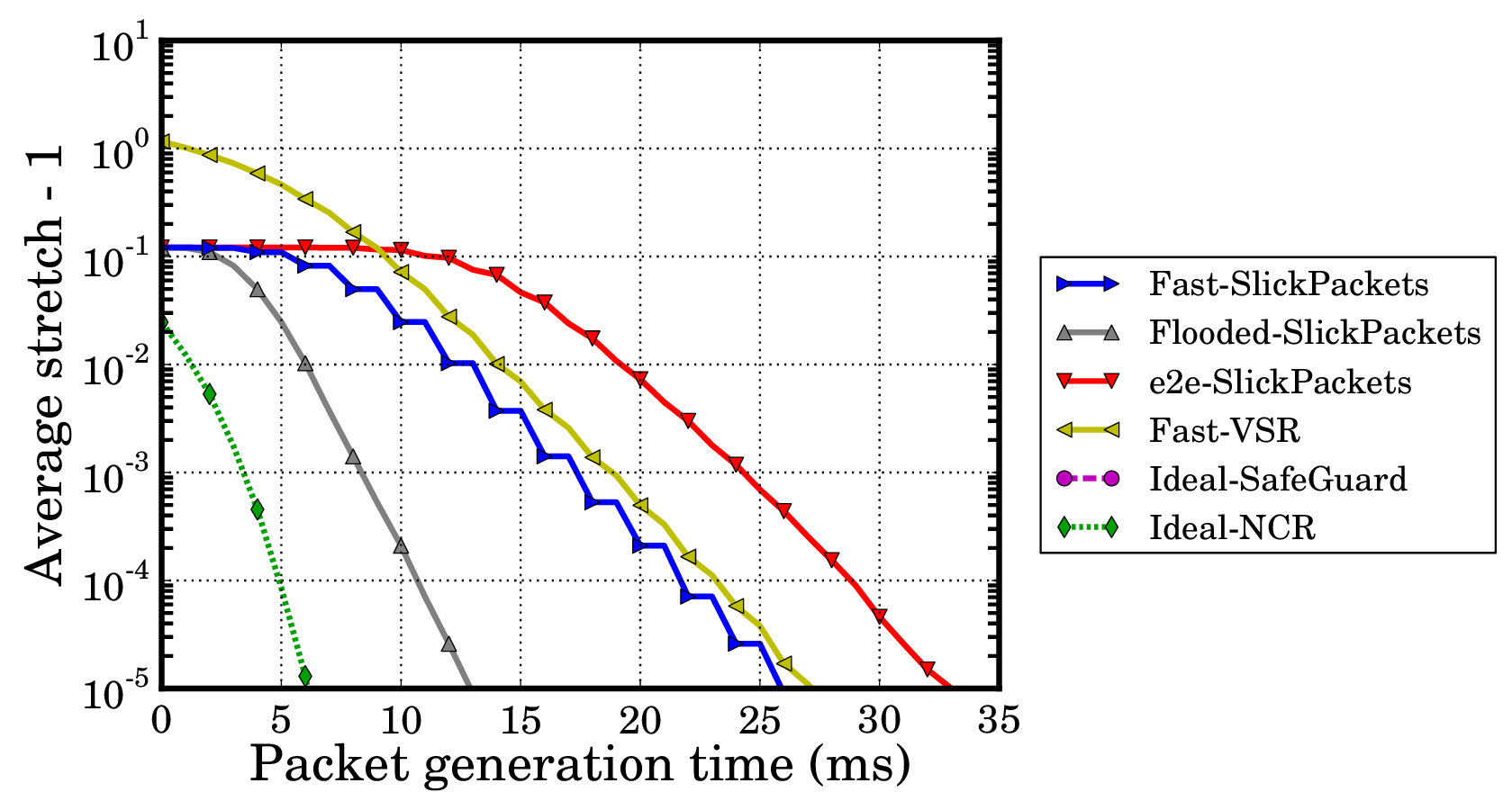}
    \label{fig:avgsvst_t0_is_0_routerlevel}
  }
  \caption{Average packet stretch - 1 vs. packet generation time when
    $t_0=0$. The y-axes are on log scales. For the Sprint
    topology, $D=50,d_r=2$. For the AS- and router-level topologies,
    $D=d_r=0$.}
  \label{fig:avg_t0_is_0}
\end{figure*}
\begin{figure*}
  \centering
  \subfigure[Sprint Topology]{
    \includegraphics[height = 1.4in]{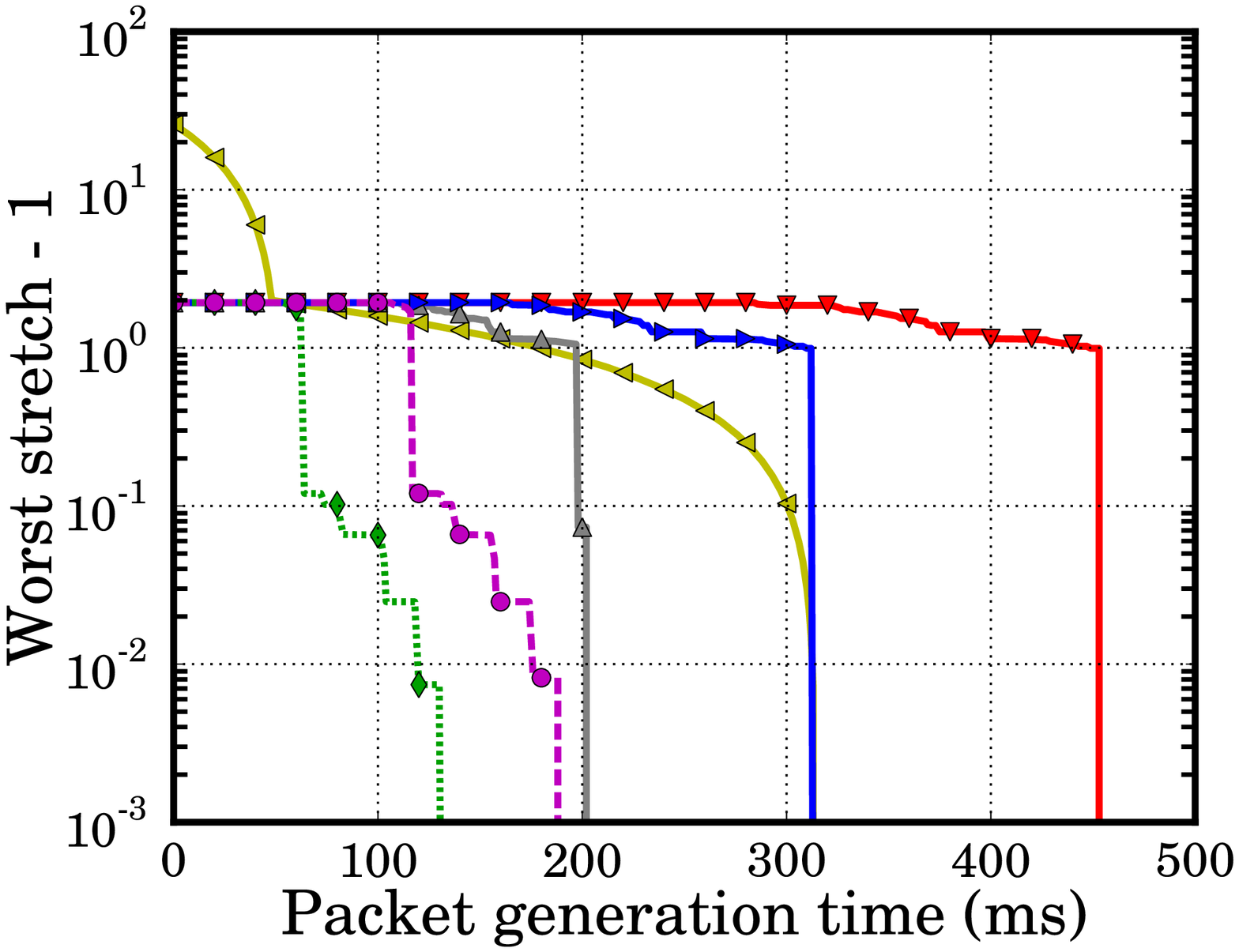}
    \label{fig:maxsvst_t0_is_0_rf1239}
  }
  \subfigure[AS-level Topology]{
    \includegraphics[height = 1.4in]{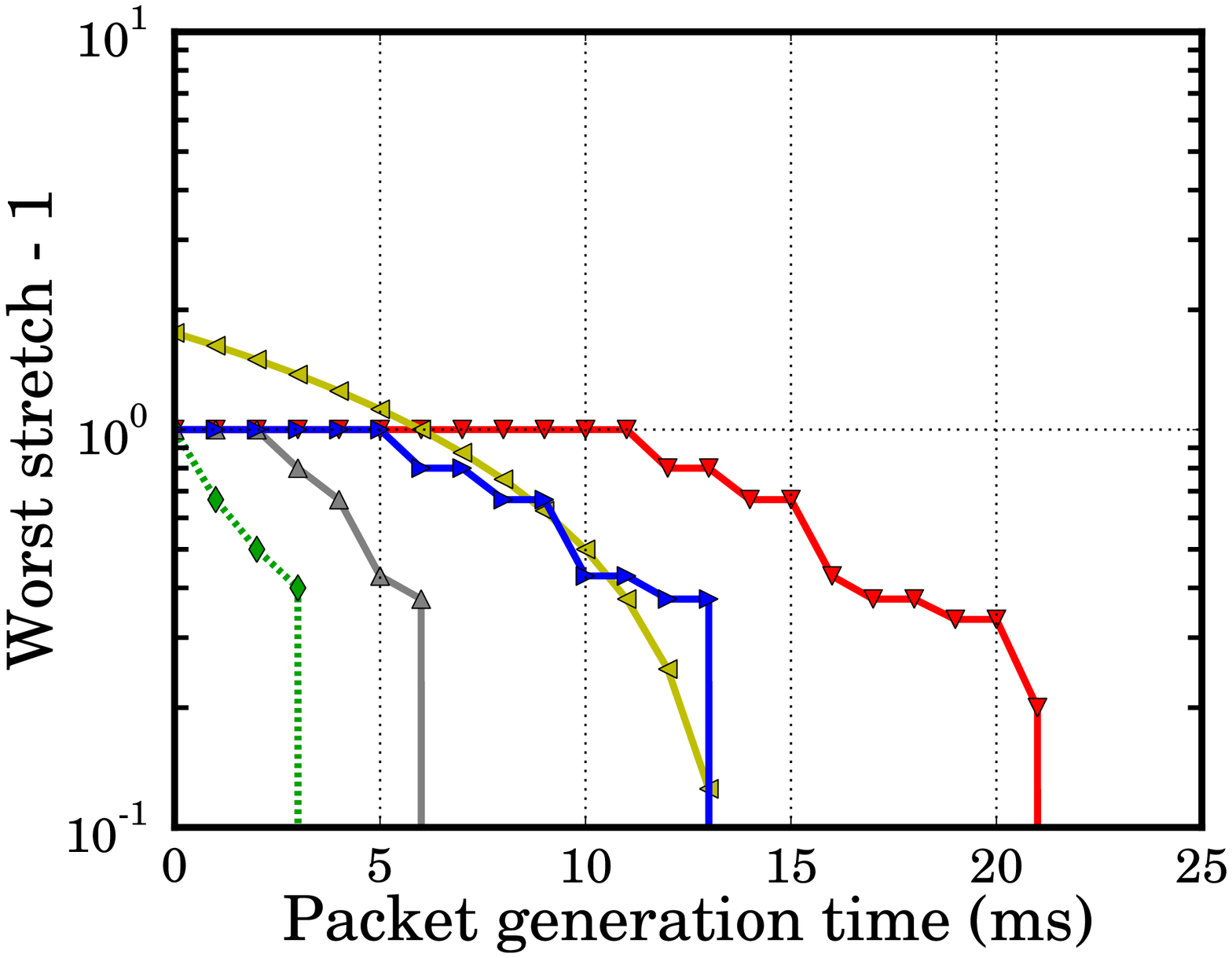}
    \label{fig:maxsvst_t0_is_0_aslevel}
  }
  \subfigure[Router-level Topology]{
    \includegraphics[height = 1.4in]{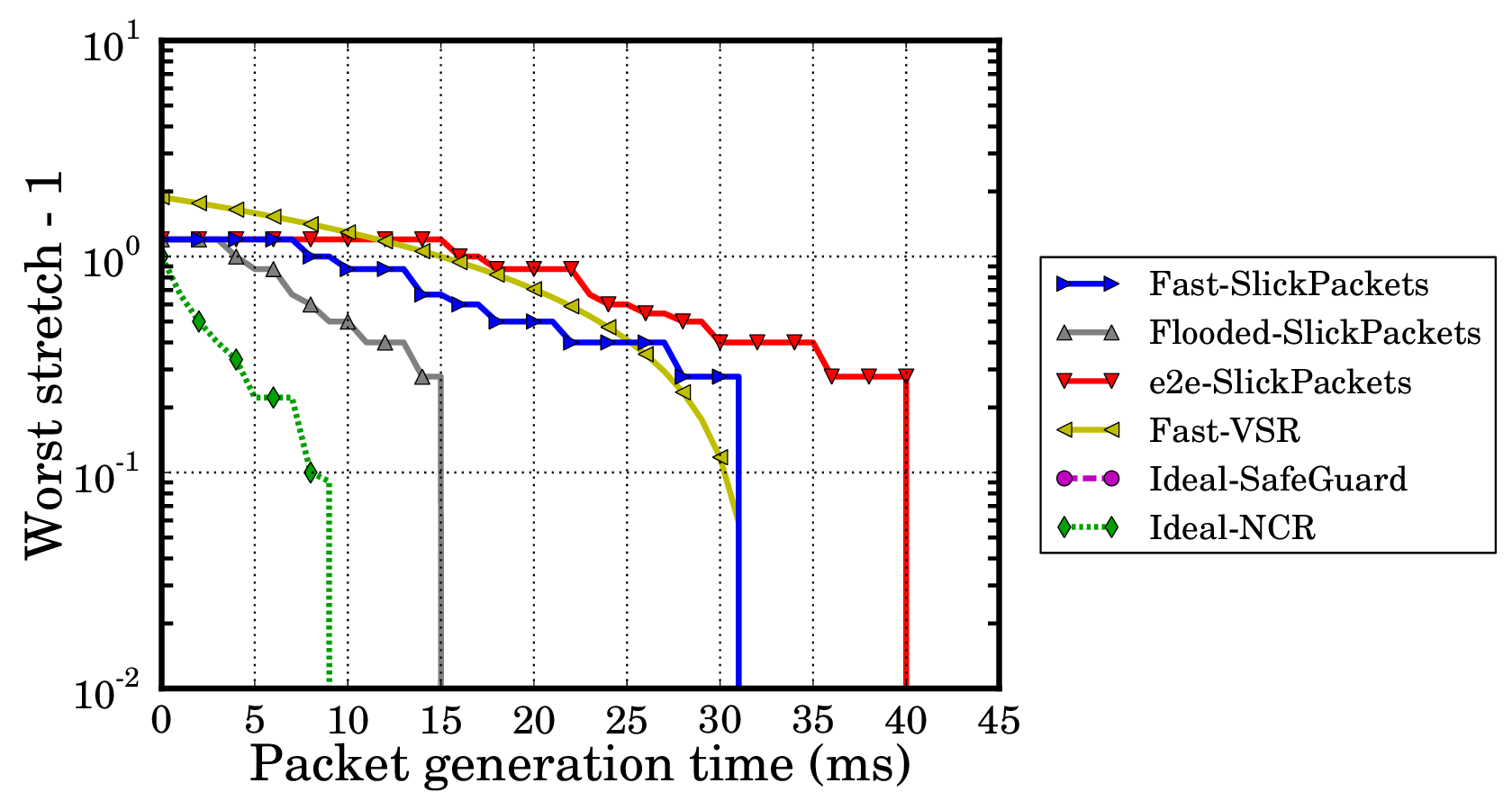}
    \label{fig:maxsvst_t0_is_0_routerlevel}
  }
  \caption{Worst packet stretch vs. packet generation time when
    $t_0=0$. The y-axes are on log scales. For the Sprint
    topology, $D=50,d_r=2$. For the AS- and router-level topologies,
    $D=d_r=0$.}
  \label{fig:worst_t0_is_0}
\end{figure*}
\subsubsection{Results}
\label{ssec:sr}
The
  high-level results reveal that \name schemes (particularly the Fast
  and Flooded variants) achieve packet stretch comparable to that of
  NCR scheme Ideal-\sg. Although \name schemes take slightly longer to
  converge compared to \sg, they avoid the high packet stretch of Fast-VSR.

\paragraphb{Average stretch}Fig. \ref{fig:avgsvst_t0_is_diameter}
shows the packet stretch averaged over all evaluated $(l_0, s, d)$
triples when $t_0$ is greater than the network diameter. We first consider features common to all
  schemes. For a given scheme, all packets generated early in the
  simulation have stretch 1. Gradually, as packets generated closer to
  $t_0$, as well as more triples where $s$ is closer to $l_0$, are
  affected by the failure, the average stretch
  increases. Additionally, for any triple, all packets generated after
  $t_0$ have stretch no higher than those generated at $t_0$; this is
  reflected in the average stretch over all triples.

  We now compare NCR and \name schemes. In NCR schemes, routers
  upstream from $l_0$, once they receive the LSA and update their
  FIBs, can redirect packets before they reach $l_0$; while in \name
  schemes, packets have to reach $l_0$ before being redirected. This
  difference gives NCR schemes only a small advantage for early
  packets, especially for the Sprint topology in
  Fig.~\ref{fig:avgsvst_t0_is_diameter}(a), because upstream routers
  still incur the delay $D$ between receiving the LSA and updating
  their FIBs. For later packets, this
  advantage becomes more significant as more upstream
  routers update their FIBs. As expected, Ideal-NCR is the best
  performing scheme in all three topologies: it converges
  57~ms before Ideal-\sg for the Sprint topology (due to $D=50$ and $d_r=2$) and
  is equivalent to Ideal-\sg (not shown) in the other two topologies,
  where $D=d_r=0$.

  Consider the \name schemes in
  Fig.~\ref{fig:avgsvst_t0_is_diameter}(a). We see that for packets
  generated between $t_0=150$ and $t_0+D=200$, the average packet
  stretch is (1) constant within the same scheme and (2) identical
  across all schemes. Recall that all \name schemes use the same FS
  selection algorithm and incur the same delay $D$ between learning of
  the failure and switching to new primary paths. Thus, the only
  factor affecting their relative performances is the time $s$ learns
  of the failure, which is determined by the relative distances among
  $l_0$, $s$, and $d$ for different triples in the same scheme, and
  the different control schemes given the same triple. So, regardless
  of the $(l_0, s, d)$ triple or the control scheme, there is a
  minimum window of $D$ time where $s$ uses the same (old) primary
  path. After this window, we can see that Fast-\name
  converges slightly faster than Flooded-\name because the LSAs in
  Flooded-\name incur delay $d_r$ at intermediate routers; in
  Fig.~\ref{fig:avgsvst_t0_is_diameter}(b) and (c), where $d_r=0$,
  Fast- and Flooded-\name are identical. And both of them converge
  significantly faster than e2e-\name as expected.

Finally, we see that in Fast-VSR, early packets experience higher
stretch than in other schemes. This is because these packets are
dropped and have to be resent by $s$. They 
experience on average a delay of one half the RTT between $s$ and $r_0$, plus the delay $D$ before being sent along the new path, resulting in a high stretch. However, Fast-VSR can catch up to and overtake Fast-\name for two reasons. First, consider the packet sent 1~ms before $s$ learns of the failure: in Fast-VSR, it is delayed $(1+D)$~ms before being resent along the new path; while in Fast-\name, the amount of time this packet traverses the original primary path only to be redirected backwards can be larger than $(1+D)$, especially if both the primary path and alternate path contain a very high latency link. Second, consider the packet generated 1~ms before $s$ has a new primary path: in Fast-VSR, it is delayed (queued) only 1~ms before being sent on the new optimal path; while in Fast-\name, this packet will be sent along the original primary path and will be redirected, experiencing a higher stretch than its Fast-VSR counterpart. These two effects enable Fast-VSR to noticeably overtake Fast-\name in Fig.~\ref{fig:avgsvst_t0_is_diameter}(a), but in Fig.~\ref{fig:avgsvst_t0_is_diameter}(b) and (c), where $D=0$ and all links have latencies 1~ms, these two effects are less pronounced.

\paragraphb{Worst stretch}Fig.~\ref{fig:worst_t0_is_diameter}
  shows the worst stretch of packets given their generation time,
  among all evaluated $(l_0, s, d)$ triples, when $t_0$ is greater
  than the network diameter. Note that the simulation-wide
  worst stretches for all schemes except Fast-VSR are equal, which are
  2.93, 2.0, and 2.2 in Fig.~\ref{fig:worst_t0_is_diameter}(a), (b),
  and (c), respectively. This is because all these schemes do not drop
  packets, so the worst stretch is that of packets that $r_0$
  redirects, which is the same for all these schemes. Also
    note that for schemes that do not drop or queue packets, the worst stretch
    occurs when a packet traverses the maximum possible distance along
    the original shortest path without reaching $d$, is redirected
    back to $s$, and traverses the shortest alternate path. So, 3 is
    the upper-bound stretch because the shortest alternate path cannot
    be shorter than the original shortest path.

  For the Sprint topology in Fig.~\ref{fig:worst_t0_is_diameter}(a),
  the simulation-wide worst stretch for Fast-VSR is 27. This happens
  to packets sent right before $t_0=150$ in triples where $s$
  is close to $d$, so that the time duration $D$ that
  these packets are delayed dominates the latencies of the original
  and post-link-failure shortest paths. In the AS- and router-level
  topologies, where $D=0$, the simulation-wide worst stretch of
  Fast-VSR are 2.75 and 2.88 respectively.

\paragraphb{When $t_0=0$} Fig. \ref{fig:avg_t0_is_0} and
  \ref{fig:worst_t0_is_0} show the results for when $t_0=0$. The
  overall behavior of each individual scheme exhibits similar patterns
  to when $t_0$ is greater than the network diameter. The differences
  are that the peak stretches occur for packets generated at
  $t_0=0$. Furthermore, as expected, flooding schemes benefit from the
  earlier time of failure: for example, for the Sprint topology in
  Fig.~\ref{fig:avg_t0_is_0}(a), Ideal-NCR and Ideal-\sg converge
  further ahead of Fast-\name compared to
  Fig.~\ref{fig:avgsvst_t0_is_diameter}(a), and even Flooded-\name now
  converges ahead of Fast-\name (similarly for the AS- and
  router-level topologies).

  In terms of simulation-wide worst stretch, those of non-flooding
  schemes (Fast- and e2e-\name as well as Fast-VSR) are the same as
  when $t_0$ is greater than the network diameter. This is as expected
  because for these schemes, it is still $r_0$ that redirects packets
  and/or triggers the notification of sources. For flooding schemes,
  however, it can be expected that simulation-wide worst stretch would
  be lower compared to when $t_0$ is greater than the network
  diameter. Nevertheless, the Sprint topology contains triples where
  an upstream link that is close to $r_0$ has very high latency
  compared to the distance between $s$ and $r_0$, so that $s$'s first
  packet does not benefit from the flooded LSA: it still has to reach
  $r_0$ before being redirected. This results in the simulation-wide
  worst stretch of 2.93 in Fig. \ref{fig:avg_t0_is_0}(a).

\section{Discussion: Forwarding\\Subgraph Selection}
\label{sec-extensions}

The \name design is agnostic to how the source selects the forwarding subgraph (FS). For example, the FS selection may be guided by demands of the application running at the source (for example, if the source is an end host) or the performance goals of a network operator (for example, if the source is an edge router). In this paper, we presented and evaluated one such FS selection algorithm: where the FS allows re-routing of packets within the network in case of single-link failures. We now discuss alternative FS selection strategies. 

\paragraphb{Handling node failures} For the FS to handle node failures, we need only a simple modification to the link-failure-avoiding FS selection of \S\ref{sec-design-1}. A source $s$, for a given destination $d$, constructs
the FS in three steps. First, $s$ computes a primary path
$P$ to $d$ by running an instance of the shortest path algorithm. Next, to
protect against single {\em node} failures, $s$ visits each node along $P$,
and computes the alternate path $P_i$ it would prefer the packet to be
routed along if that node were to fail. In particular, for each node
$v_i$ on the primary path with node $v_{i+1}$ as the next hop along the primary path, we (a) remove $v_{i+1}$; (b) compute a
shortest path from $v_i$ to $d$;
and, (c) restore $v_{i+1}$.

\paragraphb{Handling multiple link failures} A source may desire to construct
an FS that protects against multiple link failures.  This may be done by
extending the scheme from \S\ref{sec-design} to construct an FS  that
protects from multiple edge-failures.  For example, it may be sufficient to
have two strategically chosen \deto  paths for all nodes on the primary path.
The idea is that the source can choose \deto paths that are not
failure-correlated with the primary path. This may allow a  much larger amount
of resiliency; although the performance evaluation of such a scheme is
subject to future work.

\paragraphb{Congestion avoidance} Our focus in this work so far has been on
dealing with failures. However, alternate paths in the FS may also be used to
react to congestion in the network.  For example, intermediate routers along
the path may choose to forward the packet along an alternate path if the
primary path is congested (e.g., if the interface queue for the corresponding
link is filled beyond a particular threshold).  Using a FS also enables the
source to optionally provide control over load balancing, by
providing feedback on which set of paths are tolerable for the load balancing
process.

\section{Related Work}
\label{sec-relatedwork}

Our goals are related to two key areas of related work: 

\paragraphb{Failure reaction in network-controlled routing protocols} There has been
much work on coping with failures in IP networks.  We focus on the most closely
related work: protocols that guarantee packet delivery in the presence of one or more
link failures.  R-BGP~\cite{r-bgp} constructs interdomain backup paths  to
handle single link failures, given some assumptions about routing policies.
\sg~\cite{safeguard} uses a remaining path cost field in a packet as a
heuristic to determine whether the path expected by the previous hop is
different than the path available to the current hop.  In this way, it can
decide when to reroute packets along pre-computed backup paths.  FCP~\cite{fcp}
takes a different approach to determining when packets should be rerouted: each
packet carries a list of the failed links it has encountered.  The best backup
paths are computed on the fly at routers, thus allowing FCP to be robust to
multiple link failures, but requiring fairly heavyweight graph processing in
the data plane.  MPLS Fast Reroute~\cite{rfc4090} relies on precomputation of
backup paths.  In its local repair variant, an additional path is constructed
to avoid each neighboring link or node, which can inflate storage
requirements and will not result in lowest-stretch backup paths.  As discussed in the
introduction, all of the above approaches are NCR protocols,
which do not permit source control of primary or backup paths.  In addition,
backup paths are computed or stored at every router within the network, so that
there is a dependency between each router's forwarding table and the topology
of the entire network. 

One way to get a small amount of route control at the source within an NCR architecture is to use multihoming: the source can then select between several providers~\cite{akella2004comparison}.  This could be used to enable some source control, while still applying the NCR resilience techniques described above.  However, this provides only a very limited amount of control to the source, and does not yield the full benefits of source control described in the introduction.  Moreover, if many sources are multihomed, this vastly increases routing state within the network, since each router would be required to know about every point of multihoming attachment if we desire to provide alternate paths that avoid a failure of one of these links.

Our use of routing along FSes was inspired by \cite{dag}, which argues that a directed acyclic graph is a better forwarding architecture than the more traditional
shortest-path tree.  While \cite{dag} focuses on improving NCR schemes, we target achieving the benefits of both network- and source-controlled
routing.  Additionally, while \cite{dag} will deliver every packet even
during link failures, it does not guarantee the latency that these packets will
have.  \name can guarantee that for single-link failures, packets will follow
the shortest \deto path from the point of failure to the destination.

\paragraphb{Source routing} There is also a large body of work on source
controlled routing, ranging from dynamic source routing in wireless networks
\cite{johnson1996dynamic} to future interdomain routing
architectures~\cite{nira,pathlets,motiwala08,yang06}. Two of these, Routing Deflections~\cite{yang06} and Path
Splicing~\cite{motiwala08}, target fast re-routing within the network. Both use
path label bits set by the source to pseudorandomly select a next hop at each
router or AS. In~\cite{motiwala08}, pseudorandom forwarding can lead to
forwarding loops. In~\cite{yang06} routers follow certain rules that ensure
loop-freedom, but reduce path diversity.

There are three important differences between \cite{motiwala08,yang06} and \name.  First, \cite{motiwala08,yang06} do not fully support source control over primary or backup routes; although sources can select among some set of paths, they cannot tell which paths they are selecting.  Second, although packets can be rerouted quickly within the network after a link failure, this is not guaranteed (packets may be dropped), and the backup paths are not guaranteed to have optimal latency. Third, \cite{motiwala08,yang06} are similar to traditional NCR schemes in terms of the state in the network; indeed, \cite{motiwala08} increases forwarding table size because each router stores multiple next-hops for each destination. In contrast, \name enables source control, can guarantee resilience\footnote{Unless, of course, no \deto path exists.} to single-link failures with packets sent along the shortest \deto path from the point of failure to the destination, and requires only local state at routers.

Giving sources control over constructing end-to-end paths introduces
a number of practical questions, for example in terms of policy compliance,
security, and scalability of disseminating topological state. For these questions,
we rely on previous work (e.g., ~\cite{nira,pathlets}, and citations within),
which provide solutions to these problems.

\section{Conclusion}
\label{sec:conclusion}

In this paper, we presented \name, an approach to routing that attains failure
reaction, while simultaneously retaining the benefits of source routing. \name
works by compactly encoding a set of alternate paths into data packet headers
as a {\em directed acyclic graph}. Towards this goal, we provide simple
algorithms for computing efficient graphs, and for encoding them into packets
in a manner that can be  processed by intermediate routers in an efficient
manner.

One major area left for future work is to evaluate the complexity of implementing \name in production routers, and achievable packet forwarding rates; a key challenge here is dealing with increased header size. A promising avenue for evaluation is the Supercharged PlanetLab Platform~\cite{turner2007supercharging}, a network processor-based platform on which John DeHart has implemented a prototype version of \name.

This work was supported by National Science Foundation grant CNS
10-40396.



\small
\bibliographystyle{abbrv}
\bibliography{paper}

\begin{thebibliography}{10}

\bibitem{caida}
{CAIDA}'s router-level topology measurements.
\newblock
  \nolinkurl{http://www.caida.org/tools/measurement/skitter/router_topology/}.

\bibitem{rocketfuel}
Rocketfuel: An {ISP} topology mapping engine.
\newblock
  \nolinkurl{http://www.cs.washington.edu/research/networking/rocketfuel/}.

\bibitem{socialnd1}
Y.-Y. Ahn, S.~Han, H.~Kwak, S.~Moon, and H.~Jeong.
\newblock Analysis of topological characteristics of huge online social
  networking services.
\newblock In {\em Proc. {ACM} {WWW}'07}, pages 835--844, May 2007.

\bibitem{akella2004comparison}
A.~Akella, J.~Pang, B.~Maggs, S.~Seshan, and A.~Shaikh.
\newblock {A comparison of overlay routing and multihoming route control}.
\newblock {\em ACM SIGCOMM}, 34(4):93--106, 2004.

\bibitem{andersen01ron}
D.~G. Andersen, H.~Balakrishnan, M.~F. Kaashoek, and R.~Morris.
\newblock Resilient overlay networks.
\newblock In {\em Proc. 18th ACM SOSP}, October 2001.

\bibitem{drcch10}
M.~Caesar, M.~Casado, T.~Koponen, J.~Rexford, and S.~Shenker.
\newblock Dynamic route computation considered harmful.
\newblock {\em ACM SIGCOMM Computer Communication Review}, 2010.

\bibitem{clark02tussle}
D.~Clark, J.~Wroclawski, K.~Sollins, and R.~Braden.
\newblock Tussle in cyberspace: {Defining} tomorrow's {Internet}.
\newblock In {\em SIGCOMM}, 2002.

\bibitem{delay2}
P.~Francois, C.~Filsfils, J.~Evans, and O.~Bonaventure.
\newblock Achieving sub-second {IGP} convergence in large {IP} networks.
\newblock {\em SIGCOMM Computer Communications Review}, 35:35--44, 2005.

\bibitem{delay1}
J.~Fu, P.~Sjodin, and G.~Karlsson.
\newblock Intra-domain routing convergence with centralized control.
\newblock {\em Computer Networks}, 53, 2009.

\bibitem{pathlets}
P.~B. Godfrey, I.~Ganichev, S.~Shenker, and I.~Stoica.
\newblock Pathlet routing.
\newblock In {\em ACM SIGCOMM}, 2009.

\bibitem{gummadi04improving}
K.~P. Gummadi, H.~V. Madhyastha, S.~D. Gribble, H.~M. Levy, and D.~Wetherall.
\newblock Improving the reliability of {Internet} paths with one-hop source
  routing.
\newblock In {\em Proc. OSDI}, 2004.

\bibitem{vickery}
J.~Hershberger and S.~Suri.
\newblock Vickery prices and shortest paths: what is an edge worth.
\newblock In {\em IEEE FOCS}, 2001.

\bibitem{routeview}
Y.~Hyun, B.~Huffaker, D.~Andersen, E.~Aben, M.~Luckie, kc~claffy, and
  C.~Shannon.
\newblock The ipv4 routed /24 as links dataset, November 2010.
\newblock
  \nolinkurl{http://www.caida.org/data/active/ipv4_routed_topology_aslinks_dat%
aset.xml}.

\bibitem{icmbd02}
G.~Iannaccone, C.-N. Chuah, R.~Mortier, S.~Bhattacharyya, and C.~Diot.
\newblock Analysis of link failures in an {IP} backbone.
\newblock In {\em IMC}, 2002.

\bibitem{johnson1996dynamic}
D.~Johnson and D.~Maltz.
\newblock {Dynamic source routing in ad hoc wireless networks}.
\newblock {\em Mobile computing}, pages 153--181, 1996.

\bibitem{r-bgp}
N.~Kushman, S.~Kandula, D.~Katabi, and B.~Maggs.
\newblock {R-BGP}: Staying connected in a connected world.
\newblock In {\em NSDI}, 2007.

\bibitem{fcp}
K.~Lakshminarayanan, M.~Caesar, M.~Rangan, T.~Anderson, S.~Shenker, and
  I.~Stoica.
\newblock Achieving convergence-free routing using failure-carrying packets.
\newblock {\em SIGCOMM Comput. Commun. Rev.}, 37(4):241--252, 2007.

\bibitem{safeguard}
A.~Li, X.~Yang, and D.~Wetherall.
\newblock Safeguard: Safe forwarding during route changes.
\newblock In {\em Proc. ACM CoNext}, December 2009.

\bibitem{dag}
J.~Liu, J.~Rexford, M.~Schapira, S.~Shenker, and J.~Naous.
\newblock Routing along {DAGs}, 2010.
\newblock http://www.cs.berkeley.edu/$\sim$liujd/RAD.pdf.

\bibitem{motiwala08}
M.~Motiwala, M.~Elmore, N.~Feamster, and S.~Vempala.
\newblock Path splicing.
\newblock In {\em ACM SIGCOMM}, 2008.

\bibitem{ospf}
J.~Moy.
\newblock {\em {OSPF}: Anatomy of an {Internet} Routing Protocol}.
\newblock 1998.

\bibitem{rfc4090}
P.~Pan, G.~Swallow, and A.~Atlas.
\newblock Fast reroute extensions to {RSVP-TE} for {LSP} tunnels.
\newblock In {\em RFC4090}, May 2005.

\bibitem{qiu03}
L.~Qiu, Y.~R. Yang, Y.~Zhang, and S.~Shenker.
\newblock On selfish routing in {Internet}-like environments.
\newblock In {\em {Proc. ACM SIGCOMM}}, pages 151--162, 2003.

\bibitem{savage99detour}
S.~Savage, T.~Anderson, A.~Aggarwal, D.~Becker, N.~Cardwell, A.~Collins,
  E.~Hoffman, J.~Snell, A.~Vahdat, G.~Voelker, and J.~Zahorjan.
\newblock Detour: Informed {Internet} routing and transport.
\newblock In {\em IEEE Micro}, January 1999.

\bibitem{turner2007supercharging}
J.~Turner, P.~Crowley, J.~DeHart, A.~Freestone, B.~Heller, F.~Kuhns, S.~Kumar,
  J.~Lockwood, J.~Lu, M.~Wilson, C.~Wiesman, and D.~Zar.
\newblock {Supercharging planetlab: a high performance, multi-application,
  overlay network platform}.
\newblock {\em ACM SIGCOMM}, 2007.

\bibitem{1159956}
F.~Wang, Z.~M. Mao, J.~Wang, L.~Gao, and R.~Bush.
\newblock A measurement study on the impact of routing events on end-to-end
  internet path performance.
\newblock {\em SIGCOMM Comput. Commun. Rev.}, 36(4):375--386, 2006.

\bibitem{wa+06}
D.~Wendlandt, I.~Avramopoulos, D.~Andersen, and J.~Rexford.
\newblock Don't secure routing protocols, secure data delivery.
\newblock In {\em HOTNETS}, 2006.

\bibitem{xu06miro}
W.~Xu and J.~Rexford.
\newblock {MIRO: Multi-path Interdomain ROuting}.
\newblock In {\em SIGCOMM}, 2006.

\bibitem{nira}
X.~Yang, D.~Clark, and A.~Berger.
\newblock {NIRA}: a new inter-domain routing architecture.
\newblock {\em IEEE/ACM Transactions on Networking}, 15(4):775--788, 2007.

\bibitem{yang06}
X.~Yang and D.~Wetherall.
\newblock Source selectable path diversity via routing deflections.
\newblock In {\em ACM SIGCOMM}, 2006.

\end{thebibliography}
%
\appendix
\normalsize
\section{Computing Packet Stretch}
\label{sec-app-computestretch}

\begin{table*}[ht]
\centering
\begin{tabular}{|l|l|}
\hline
Notation & Description\\
\hline
\hline
$l_0$ & a failed link on the primary path from source $s$ to
destination $d$ \\
\hline
$r_0$ & the node upstream from and adjacent to $l_0$ ($r_0$ might be source $s$ itself)\\
\hline
$t_0$ & the time of failure of link $l_0$ \\
\hline
$t_{learn}(u)$ & the time a node $u$ learns of the failure, by
detecting it or being notified ($t_{learn}(u) \geq t_{0}$)\\
\hline
$\tau(u)$ & the time a node $u$ can start sending packets along the
post-failure shortest path from $u$ to $d$\\
\hline
$\sigma(u)$ & the stretch experienced by packets that a node $u$ sends along the post-failure shortest path from $u$ to $d$\\
\hline
$\dist(u,v)$ & the shortest path latency between nodes $u$ and $v$
with no link failure\\
\hline
$\widehat{\dist}(u,v,l)$ & the shortest path latency between nodes $u$ and $v$
after the failure of link $l$\\
\hline
$\hopcount(u,v)$ & the hop-count between nodes $u$ and $v$ with no link failure\\
\hline
\end{tabular}
\caption{Notation used in the packet stretch computations}
\label{tab:stretchvstimenotations}
\end{table*}

We describe here the stretch computations our simulator
in \S\ref{ssec-simulator} uses. Given a source-destination pair $s,d$
and a failed link $l_0$ on the primary path between $s$ and $d$, we
wish to compute the stretch experienced by each packet that the
application at source $s$ generates. We assume the application at $s$
generates packets every 1ms, starting at time $t=0$. Further, we
assume that all nodes in the network have sufficient queue space so
that no packet is dropped for lack of queue space, and that the nodes
can fully flush their queues/buffers instantaneously. We assume links
have sufficient capacity and devices have sufficient data-plane
processing capabilities, so that they do not introduce delays to data
packets.

\begin{figure}[h]
		\centering
		\begin{tikzpicture}[xscale=0.43, yscale=0.43]

    		\tikzstyle{sd}=[circle,draw,fill=white,inner sep=1pt];
    		\tikzstyle{pnode}=[circle,draw,fill=blue!20,inner sep=1pt,minimum width=.5cm];
    		\tikzstyle{snode}=[circle,draw,fill=violet,inner sep=1pt];
    		\tikzstyle{vnode}=[circle,draw,fill=red,inner sep=1pt];
    		\tikzstyle{ellipsis}=[circle,draw,fill=white!20,inner sep=1pt];

     		\begin{scope}[xshift=4.25cm,>=latex]
				\node[thick] (s) at (0, 0) [pnode] {$s$};
				\node[thick] (ellipsis1) at (2.5, 0) {...};
      			\node[thick] (r1) at (5, 0) [pnode] {$r_1$};
      			\node[thick] (r0) at (7.5, 0) [pnode] {$r_0$};
				\node[thick] (ellipsis2) at (10, 0) {...};
      			\node[thick] (d) at  (12.5, 0) [pnode] {$d$};

				\node[thick, red] at (8.75, 0) {X};
				\node[thick, red] at (8.85, 0.8) {$l_0$};

                \draw[thick] (s) -- (ellipsis1);
				\draw[thick] (ellipsis1) -- (r1);
				\draw[thick] (r1) -- (r0);
				\draw[thick] (r0) -- (ellipsis2);
				\draw[thick] (ellipsis2) -- (d);
     		\end{scope}
  		\end{tikzpicture}
\caption{A primary path from $s$ to $d$ with failed link $l_0$. $r_0$ is the router upstream from and adjacent to $l_0$.}
\label{fig:toy2}
\end{figure}
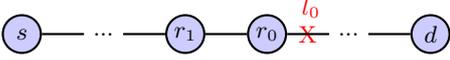

First, we give an overview of our approach (notation summarized in
Table~\ref{tab:stretchvstimenotations}). For a given $(l_0,s,d)$
triple in Fig.~\ref{fig:toy2}, consider a router $r$ on the primary path that is upstream
from $l_0$. After the failure of $l_0$, router $r$ can
``offer'' two types of stretch to packets that reach it: (a) to
packets that $r$ redirects along its alternate path to $d$, a fixed
stretch $\sigma(r)$; (b) to packets that $r$ forwards along its primary
path, whatever stretch offered by downstream routers on the primary
path. The two important
features of a router $r$ are then the time $\tau(r)$ at which
it starts redirecting packets along its alternate path, and the fixed
stretch $\sigma(r)$ it offers such redirected packets. Then, given
$\tau(r)$, we can compute the sent time of the first packet from $s$
that will be redirected by $r$; packets sent before this packet will
be forwarded to $r$'s primary next-hop and thus experience whatever
stretch offered by $r$'s downstream routers. Applying the above
analysis to all routers upstream from $l_0$, our simulator determines
the stretch experienced by any packet given the time it is generated
by the source application.

\subsection{\nametitle}
With all \name variants, for a particular $(l_0,s,d)$ triple, the only
two nodes that potentially redirect packets are $r_0$ and source $s$,
so we only need to analyse these two nodes.

\paragraphb{Consider $r_0$} It starts redirecting packets at time
$\tau(r_0)=t_0$. The first packet from $s$ that will be redirected by $r_0$
arrives at $r_0$ at time $\max\{t_0,\dist(s,r_0)\}$ and thus is
sent/generated by $s$ at time
\begin{align}
\max\{t_0, \dist(s, r_0)\} - \dist(s,r_0)
\end{align}
This packet and all packets redirected by $r_0$ experience its fixed stretch, which is
\begin{align}
\sigma(r_0)=\dfrac{\dist(s,r_0)+\widehat{\dist}(r_0,d,l_0)}{\widehat{\dist}(s,d,l_0)}
\end{align}

\paragraphb{Consider $s$} It is easy to see that
\begin{align}
\sigma(s)=1
\end{align}
To compute the time that $s$ first ``redirects'' packets---sends them
along its new primary path to $d$---note that for all \name variants,
this time is
\begin{align}
\tau(s)=t_{learn}(s)+D\nonumber
\end{align}
Now, we show the derivation of $t_{learn}(s)$ for each \name variant.

\paragraphb{Flooded-\name} At time $t_0$, when it detects the link failure---not when it receives packets from source $s$---the router $r_0$ floods the network with the LSA. Thus, $s$ receives the LSA and learns of the link failure at time
\begin{align}
t_{learn}(s)=t_0 + \dist(r_0,s) + \hopcount(r_0, s) \cdot d_r
\end{align}

\paragraphb{Fast-\name} Upon receiving a packet from $s$ that requires
redirection, $r_0$ sends a notification message to $s$ about the
failed link. Thus, $s$ receives the notification message and learns of the link failure
at time
\begin{align}
t_{learn}(s)=\max\{t_0,\dist(s,r_0)\}+\dist(s,r_0)
\end{align}

\paragraphb{e2e-\name} Upon receiving a packet from $s$ that requires
redirection, $r_0$ piggybacks the notification message in the packet
on the way to $d$, which receives the packet at time
$\max\{t_0,\dist(s,r_0)\}+\widehat{\dist}(r_0,d,l_0)$. After $d$ computes its
new shortest path to $s$, it sends the notification message to
$s$. Thus, $s$ receives the notification message and learns of the
link failure at
time
\begin{align}
t_{learn}(s)=&\max\{t_0,\dist(s,r_0)\}+\widehat{\dist}(r_0,d,l_0)\nonumber\\
&+D+\widehat{\dist}(d,s,l_0)
\end{align}

\subsection{Ideal-\sg} At time $t_0$, when it detects the link failure---not when it receives packets from $s$---the router $r_0$ floods the network with the LSA. As routers upstream from $r_0$ learn about the link failure, they can potentially redirect packets along shorter paths, reducing the amount of time packets traverse a sub-optimal path, i.e., packets do not have to reach the failure before being redirected.

\paragraphb{Consider $r_0$} It learns of the link failure at time $t_0$. Because it has
already pre-computed alternate paths to $d$, it can immediately
redirect packets along the alternate path. Thus the first packet from
$s$ that will be redirected by $r_0$ arrives at $r_0$ at time
$\max\{t_0,\dist(s,r_0)\}$ and thus is sent/generated by $s$ at time
\begin{equation}
\max\{t_0, \dist(s, r_0)\} - \dist(s, r_0)
\end{equation}
This packet and all packets redirected by $r_0$ experience its fixed
stretch
\begin{align}
\sigma(r_0)=\dfrac{\dist(s,r_0)+\widehat{\dist}(r_0,d,l_0)}{\widehat{\dist}(s,d,l_0)}
\end{align}

\paragraphb{Consider a node $r$ upstream from $r_0$} ($r$ can
be a router or source $s$.) It learns of the link failure at
time $$t_{learn}(r) = t_0 + \dist(r_0,r) + \hopcount(r_0,r) \cdot
d_r$$ Because it is not adjacent to the failed link, $r$ only starts
redirecting packets along its alternate path to $d$ at time
$t_{learn}(r)+D$. Thus, the first packet from $s$ that will be
redirected by $r$ along the alternate path is sent by $s$ at time
\begin{equation}
\max\{0, t_{learn}(r)+D-\dist(s,r)\}
\label{equ:sg-txtimeoffirstredirected}
\end{equation}
This packet and all packets redirected by $r$ experience its fixed
stretch
\begin{align}
\sigma(r)=\dfrac{\dist(s, r) + \widehat{\dist}(r, d, l_0)}{\widehat{\dist}(s, d,l_0)}
\end{align}

Next, consider two adjacent routers $r$ and $r'$ upstream from $l_0$
where $r' \neq r_0$ and $r$ is upstream from $r'$. Let $t$ and $t'$, given by Eq.~\ref{equ:sg-txtimeoffirstredirected}, be the sent times of the first packets from $s$ that are
redirected by $r$ and $r'$, respectively. Note that $\dist(r_0, r)
> \dist(r_0, r')$, $\hopcount(r_0, r) > \hopcount(r_0, r')$, and
$\dist(s, r) < \dist(s, r')$; thus $t\geq t'$ ($t=t'\Leftrightarrow
t=t'=0$). If $t>t'$, packets sent by $s$ in the interval $[t',t)$ are redirected by $r'$ and thus experience the stretch offered by
$r'$.

\subsection{Fast-VSR}

Unlike \name and \sg, with VSR, the router $r_0$ drops instead of
redirecting packets; source $s$ has to resend those dropped
packets. Also, the only node that ``redirects'' packets is source
$s$---it sends them along a new primary path. Furthermore, $s$ queues
packets generated by the application between the times it learns of
the failure and is ready to use a new path to $d$. Thus, we only
consider source $s$, but we consider two types of packets: those that
are dropped and resent, and those that are queued.

The first packet from $s$ that will arrive at $r_0$ after the link
failure---thus will be dropped by $r_0$ and later resent by
$s$---arrives at time $\max\{t_0,\dist(s,r_0)\}$ (and thus is
originally sent by $s$ at time
$\max\{t_0,\dist(s,r_0)\}-\dist(s,r_0)$). Upon receiving this packet, $r_0$
sends a notification message to $s$, which $s$ receives at time
\begin{equation}
\label{equ:vsr-learn}
t_{learn}(s)=\max\{t_0,\dist(s,r_0)\} + \dist(s,r_0)\nonumber
\end{equation}
Thus, $s$ is ready to use a new primary path to $d$ at time
\begin{equation}
\label{equ:vsr-newpath}
\tau(s)=\max\{t_0,\dist(s,r_0)\}+\dist(s,r_0)+D\nonumber
\end{equation}
At this time, $s$ instantaneously resends all packets that would have
been dropped. These packets were originally sent at
time
\begin{align}
\max\{t_0,\dist(s,r_0)\}-\dist(s,r_0)+\Delta
\end{align}
with $\Delta\in[0,2\cdot\dist(s,r_0))$ (recall ``$[t-R,t)$'' in~\S\ref{ssec-schemes}) and thus experience stretch
\begin{equation}
\label{equ:vsr-resentstretch}
\dfrac{2\cdot\dist(s,r_0)-\Delta+D+\widehat{\dist}(s,d,l_0)}{\widehat{\dist}(s,d,l_0)}
\end{equation}

For the queued packets, after learning of the failure, $s$ queues all
packets newly generated by the source application until it is ready to
use a new primary path. These packets are generated between
$t_{learn}(s)$ and $\tau(s)$. In other words, they are generated at
time
\begin{align}
t_{learn}(s) + \Delta
\end{align}
with $\Delta\in[0,D]$. Since $s$ also instantaneously sends all these
queued packets at $\tau(s)$, these packets experience stretch
\begin{equation}
\label{equ:vsr-queuedstretch}
\dfrac{D-\Delta+\widehat{\dist}(s,d,l_0)}{\widehat{\dist}(s,d,l_0)}
\end{equation}
Note that the packet generated at time $t_{learn}(s)+D$ has stretch 1,
as is expected.

\section{Direct Encoding}
\label{sec-app-directcodec}

Direct Encoding embeds the Forwarding Subgraph (FS) as a directed acyclic graph data structure in the packet header. At a high level, each
router in the FS---except the destination router because it has no
outgoing links---is encoded exactly once in a structure we call
the \textbf{NodeDescriptor} (ND), at some location (bit offset) within
the encoding. A router's \textbf{NodeDescriptor} (ND)
contains \textbf{SuccessorDescriptor} (SD) structures, which represent
the router's next-hop successor(s). A \textbf{SuccessorDescriptor}
(SD) contains (1) the router's \emph{locally unique link identifier}
for the next-hop successor and (2) the offset pointer to the
successor's ND. Finally, the packet header contains a ``current node
offset pointer''. A router reads this pointer to locate its ND, and updates
it to point to the next hop's offset pointer before forwarding.

We use the overall format:


\begin{figure}[h]
\centering
\begin{tikzpicture}[yscale=0.7]
\draw[preaction={fill=black,opacity=.5, transform canvas={xshift=1mm,yshift=-1mm}}][fill=white] (0,0) rectangle (8,1); 
\draw (1.2, 0.5) node {NodePtrLength};
\draw (2.4, 0) -- (2.4, 1);
\draw (3.7, 0.5) node {CurrentNodePtr};
\draw (5.0, 0) -- (5.0, 1);
\draw (5.4, 0.5) node {$ND_1$};
\draw (5.8, 0) -- (5.8, 1);
\draw (6.2, 0.5) node {$ND_2$};
\draw (6.6, 0) -- (6.6, 1);
\draw (6.9, 0.5) node {$\dots$};
\draw (7.2, 0) -- (7.2, 1);
\draw (7.6, 0.5) node {$ND_k$};
\end{tikzpicture}
\end{figure}

\begin{description*}
\item[NodePtrLength]
A prefix code that indicates the length in bits of
the \textbf{CurrentNodePtr} field and all other absolute node
pointers. The mappings are 0, 10, 110, and 1110, for 10, 8, 6, and 4
bits respectively.

\item[CurrentNodePtr]
This value specifies the bit offset (from the beginning of the encoding)
of the current router's ND. The value of zero has a special meaning: the
current router is the final destination/egress router.
\end{description*}

An ND can have either one or two SD's, with the convention that the
first successor is the primary one. The ND has the
following format:

\begin{figure}[h]
\centering
\begin{tikzpicture}[yscale=0.7]
\draw[preaction={fill=black,opacity=.5, transform canvas={xshift=1mm,yshift=-1mm}}][fill=white] (0,0) rectangle (5.6,1); 
\draw (2.0, 0.5) node {NumberOfSuccessors};
\draw (4.0, 0) -- (4.0, 1);
\draw (4.4, 0.5) node {$SD_1$};
\draw (4.8, 0) -- (4.8, 1);
\draw (5.2, 0.5) node {$SD_2$};
\end{tikzpicture}
\end{figure}

\begin{description*}
\item[NumberOfSuccessors]
(1 bit) 0 indicates there is one successor, and 1 indicates there are
two successors.
\end{description*}

The SD contains two main pieces of information: the next-hop
identifier and the offset pointer to its ND. For the next-hop
identifier, similar to the encoding scheme discussed
in \S\ref{sec-design-2}, we use the router's \emph{locally unique link
identifiers}, which it advertises as part of the network map
dissemination. For the offset to the next-hop's ND, we use a 1-bit
flag to indicate that the next-hop's ND immediately follows the
current ND; otherwise, we include an absolute offset pointer to the
next-hop's ND. Here is the SD format:

\begin{figure}[h]
\centering
\begin{tikzpicture}[yscale=0.7]
\draw[preaction={fill=black,opacity=.5, transform canvas={xshift=1mm,yshift=-1mm}}][fill=white] (0,0) rectangle (5.6,1); 
\draw (1.0, 0.5) node {LinkId};
\draw (2.0, 0) -- (2.0, 1);
\draw (3.2, 0.5) node {ContainsPtr?};
\draw (4.4, 0) -- (4.4, 1);
\draw (5.0, 0.5) node {Ptr};
\end{tikzpicture}
\end{figure}

\begin{description*}
\item[LinkId]
The identifier of the link to forward the packet. The length of this field is
specified by the router as part of the map dissemination. In our encoding size evaluation (\S\ref{ssec-es}), we assume that it is
$\lceil\log_2\Delta\rceil$ bits.

\item[ContainsPtr?]
(1 bit) 0 indicates that the next-hop's ND follows immediately after
the current router's ND.

\item[Ptr]
Pointer to the next-hop's ND. The Length of this field is specified by
\textbf{NodePtrLength} discussed above.
\end{description*}

The 1-bit flag is only an optimization that allows us to leave off the
offset to the next-hop's ND. To make this optimization useful, the
encoding algorithm first encodes all nodes on the primary path one
after another. The first (primary) SD of each of these uses the
1-bit flag because the successor's ND immediately follows its own
descriptor (except in the penultimate router's case, which uses an
absolute pointer value of zero). The second SD, if any, uses the
absolute offset pointer.

After encoding all nodes on the primary path, the encoding algorithm
picks one of the alternate paths and encodes all of its
yet-to-be-encoded nodes one after another. These nodes that are
encoded contiguously can use the relative pointer for their SD's, and
when a node's next-hop successor is an already-encoded node, then the
next-hop successor's offset pointer is used. Also, the penultimate
router uses an absolute pointer value of zero in its SD.

\paragraphb{Forwarding Algorithm}
Upon receiving a packet, the router first gets the value of
the \textbf{CurrentNodePtr} (after parsing \textbf{NodePtrLength}). If
the value is zero, then the router is the egress router, and it can
perform appropriate actions on the packet (e.g., delivering it on
attached networks), and it does not forward the packet further.

If \textbf{CurrentNodePtr} has a non-zero offset value, then the
router parses its ND at that offset. Note, the router expects that the
lengths of its \textbf{LinkId} fields are what it advertised (e.g.,
$\lceil\log_2\Delta\rceil$ bits). With that information, the router
can fully parse its ND. If the link labeled in the first SD is online,
then the router will use that link to forward the packet. Otherwise,
if there is a second SD and its link is online, then the
router will use that link to forward the packet. Otherwise, the router
drops the packet.

Before forwarding the packet, the router needs to update
the \textbf{CurrentNodePtr}. If the used SD contains an absolute
offset pointer (i.e., its \textbf{ContainsPtr?} flag is 1), then the
router updates \textbf{CurrentNodePtr} with the value in the
SD's \textbf{Ptr}:
\[ \textbf{CurrentNodePtr} \leftarrow SD.Ptr \]
Otherwise, the successor's ND follows immediately after the current
router's ND, so to obtain the successor's ND offset, the router adds
the total length of its own ND to its own (\textbf{CurrentNodePtr})
offset, and then updates \textbf{CurrentNodePtr} with that value:
\[ \textbf{CurrentNodePtr} \leftarrow |ND|+\textbf{CurrentNodePtr} \]

\section{Sampling algorithm for simulation}
\label{sec-app-samplingalgo}

For each sampled link $l_0$, we evaluate ``qualified sources'': those
whose shortest path tree (SPT) includes $l_0$.  To find qualified sources, we sample up to
2,000 random sources and use the first 100 qualified sources, or fewer
if we find fewer qualified sources. For each qualified source $s$, we
randomly sample 100 destinations from among all those on the subtree
of $s$'s SPT that uses $l_0$. Finally, among the sampled destinations,
we use only those that remain connected with $s$ after removing $l_0$.

\section{Lower bound on edge-set size}
\label{sec-app-lowerboundalgo}
How much can we reduce the size of the FS by designing more sophisticated algorithms for selecting the FS? How close are the results given in \S\ref{sec-eval} to the smallest possible header for handling single link failures? 

In order to be able to answer the above questions, we derived lower bounds on the edge-set size of FSs that provide fast failure reaction against single link failures. That is, for any FS that uses the shortest path between the source and the destination as the primary path, the lower bound gives the minimum number of edges that the FS must contain in order to provide an alternate path avoiding any single-link failure on the primary path. For any source-destination pair $s, d$, the lower bound is given as follows:
\begin{equation*}
\left\{
\begin{array}{ll}
2 |P(s, d)| + 1 & \text{if graph weighted}\\
\\
\left\lceil \dfrac{5 |P(s, d)|}{2} \right\rceil & \text{if graph unweighted}
\end{array} \right.
\end{equation*}
where $P(s, d)$ is the primary path and $|P(s, d)|$ the number of edges in $P(s, d)$. These lower bounds impose a fundamental limitation on the header size of \name; intuitively, it is hard to reduce the header size (in bytes) significantly without reducing the edge-set size of the resulting FS. We prove the lower bound below. 

Note that a trivial lower bound on the size of the FS is $2 |P(s,d)|$ because each node in the primary must have two outgoing edges in order to provide fast failure reaction against single link failures. Theorem~\ref{thm:lbw} essentially states this bound along with an example graph demonstrating that the bound is tight.  However, if the graph is unweighted (all edges have the same weight), we can provide a better bound: intuitively, the alternate paths must include extra edges in order to ensure that they are at least as long as the primary (which is by definition the shortest). We give this improved bound in Theorem~\ref{thm:lbu}. We assume, in the following proofs, that the graph is not a multigraph and is $2$-connected.

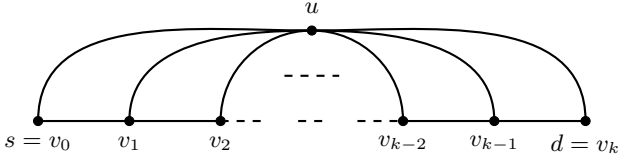
\begin{figure}[h]
\centering
    \begin{tikzpicture}[scale=0.6]
	\draw[fill=black] (0,0)  circle (1mm);
	\draw[fill=black] (2, 0)  circle (1mm); 
	\draw[fill=black] (4, 0)  circle (1mm);
	\draw[fill=black] (8, 0)  circle (1mm);
	\draw[fill=black] (10, 0)  circle (1mm);
	\draw[fill=black] (12, 0)  circle (1mm);
	\node at (0, -0.5) {$s = v_0$};
	\node at (2, -0.5) {$v_1$};
	\node at (4, -0.5) {$v_2$};
	\node at (8, -0.5) {$v_{k-2}$};
	\node at (10, -0.5) {$v_{k-1}$};
	\node at (12, -0.5) {$d = v_k$};
	\node at (6, 2.5) {$u$};

 	\draw[fill=black] (6, 2)  circle (1mm);
	
	\draw[thick] (0, 0) -- (2, 0);
	\draw[thick] (2, 0) -- (4, 0);
	\draw[dashed, thick] (4, 0) -- (5,0);
	\draw[dashed, thick] (7, 0) -- (8,0);
	\draw[dashed, thick] (5.7, 0) -- (6.3,0);
	\draw[thick] (8, 0) -- (10, 0);
	\draw[thick] (10, 0) -- (12, 0);
	\draw[thick] (0, 0) to [out=90,in=180] (6, 2);
	\draw[thick] (2, 0) to [out=90,in=180] (6, 2);
	\draw[thick] (4, 0) to [out=90,in=180] (6, 2);
	\draw[thick] (8, 0) to [out=90,in=0] (6, 2);
	\draw[thick] (10, 0) to [out=90,in=0] (6, 2);
	\draw[thick] (12, 0) to [out=90,in=0] (6, 2);
	\draw[dashed, thick] (5.4, 1) -- (6.6,1);
    \end{tikzpicture}
\caption{A graph that achieves the lower bound on the size of the \FS for weighted graphs. The weight of edges $(v_i, v_{i+1})$ are all $1$; the weight of edges $(v_{i}, u)$ is set to $k-i$.}
\label{fig:lbw}
\end{figure}

\begin{theorem}
\label{thm:lbw}
Suppose the FS uses the shortest path $P(s, d)$ as the primary path and can avoid any single link failure along the primary path. Then the \FS has at least $2 |P(s, d)| + 1$ edges. Moreover, there exist graphs for which this bound is tight.
\end{theorem}
\begin{proof}
	For weighted graphs, we note that $\FS$ contains $|P(s, d)|$ edges along the shortest path. Furthermore, each node along the shortest path requires at least one additional outgoing/incoming edge in order to provide fast reaction against single link failures. The proof follows by noting that there are exactly $|P(s,d)|+1$ nodes along the shortest path. To prove tightness of the bound, we use the graph shown in Fig.~\ref{fig:lbw}.
\end{proof}

Before going to the lower bound proof for unweighted graphs, we give some definitions to make the discussion more succinct. Let $G = (V, E)$ be the graph and given a pair of vertices $s, d$, let $FS$ be the optimal FS, meaning it has the minimum possible number of edges while satisfying the conditions in the theorem. Denote the shortest path between $s$ and $d$ as $$P = P(s,d) = (s = v_0, v_1, \dots, v_{k-2}, v_{k-1}, v_{k} = d)$$ and let $|P|$ be the number of edges in $P$. Let $G' = (V, E')$ be a densest graph (with maximum possible number of edges) such that $P$ is also the shortest between $s$ and $d$ in $G'$ and let $FS'$ be the optimal forwarding subgraph between $s$ and $d$ in $G'$.  Let $|FS|$ and $|FS'|$ be the edge-set size of the optimal forwarding subgraphs $FS$ and $FS'$. Let $Q(u, v)$ denote the shortest alternate path (as computed in \S\ref{sec-design}) between any pair of nodes $u$ and $v$ and $N(u)$ be the set of neighbors of any node $u$. 

\begin{figure}[h]
\centering
    \begin{tikzpicture}[scale=0.5]
	\draw[fill=black] (0,0)  circle (1mm);
	\draw[fill=black] (2, 0)  circle (1mm); 
	\draw[fill=black] (4, 0)  circle (1mm);
	\draw[fill=black] (6, 0)  circle (1mm);
	\draw[fill=black] (10, 0)  circle (1mm);
	\draw[fill=black] (12, 0)  circle (1mm);
	\draw[fill=black] (14, 0)  circle (1mm);
	\draw[fill=black] (16, 0)  circle (1mm);
	\node at (0, -0.5) {$s$};
	\node at (16, -0.5) {$d$};

	\draw[fill=black] (2, 2)  circle (1mm);
	\draw[fill=black] (6, -2)  circle (1mm);
	\draw[fill=black] (10, -2)  circle (1mm);
	\draw[fill=black] (14, 2)  circle (1mm);

	\draw[thick] (0, 0) -- (2, 0);
	\draw[thick] (2, 0) -- (4, 0);
	\draw[thick] (4, 0) -- (6, 0);
	\draw[thick] (10, 0) -- (12, 0);
	\draw[thick] (12, 0) -- (14, 0);
	\draw[thick] (14, 0) -- (16, 0);

	\draw[thick] (0, 0) -- (2, 2);
	\draw[thick] (2, 0) -- (2, 2);
	\draw[thick] (4, 0) -- (2, 2);
	\draw[thick] (4, 0) -- (6, -2);
	\draw[thick] (6, 0) -- (6, -2);
	\draw[thick] (12, 0) -- (14, 2);
	\draw[thick] (14, 0) -- (14, 2);
	\draw[thick] (16, 0) -- (14, 2);
	\draw[thick] (10, 0) -- (10, -2);
	\draw[thick] (12, 0) -- (10, -2);

	\draw[dashed, thick] (6.3, 0) -- (9.7, 0);
	\draw[dashed, thick] (6, -2) -- (7, -1);
	\draw[dashed, thick] (10, -2) -- (9, -1);
	\draw[dashed, thick] (7, -2) -- (9, -2);
    \end{tikzpicture}
\caption{A graph that achieves the lower bound on the size of the \FS for unweighted graphs.}
\label{fig:lbu}
\end{figure}
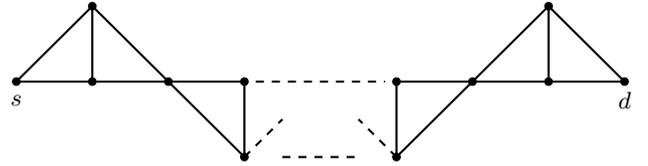

\begin{theorem}
\label{thm:lbu} 
Under the same conditions as Theorem~\ref{thm:lbw} except that edges have equal weights, the number of edges in the FS is lower bounded by: $$\left\lceil \dfrac{5 |P(s, d)|}{2} \right\rceil$$

Moreover, there exist graphs for which the bound is tight.
\end{theorem}
\begin{proof}
	We start with a few simple observations: first, since $E \subseteq E'$, we have that $|FS| \geq |FS'|$. Hence, a lower bound on $|FS'|$ implies a lower bound on $|FS|$. Second, since $G'$ is unweighted, there is no edge between $v_i$ and $v_k$ for any $i < k-1$ otherwise $P$ cannot be the shortest path. Furthermore, to provide fast failure reaction against single link failures, $FS'$ must contain an edge $(u, v_k)$ for some $u \notin P$ since $(v_{k-1}, v_k) \in P$ and the graph is not a multigraph.

Consider nodes $u$ and $v_{k-1}$. Since the graph is not a multigraph, we have that $|Q(v_{k-1}, v_k)| \geq 2$. Hence, we can replace $Q(v_{k-1}, v_k)$ by $(v_{k-1}, u, v_k)$ without increasing $|FS'|$; the edge $(u, v_k)$ indeed exists as argued earlier. Now, consider node $v_{k-2}$. Note that $|Q(v_{k-2}, v_k)| \geq 2$ and hence, we can replace $Q(v_{k-2}, v_{k})$ by $(v_{k-2}, u, v_k)$ as earlier. 

We make a final observation: let us denote by $FS' \backslash P$ the set of nodes that are in the $FS'$ but not in $P$. We claim that for any node $q \in FS' \backslash P$, $N(q) \cap P \leq 3$. To prove this, suppose by way of contradiction that $N(q) \cap P \geq 4$. Then at least two of the nodes in $N(q)$ are at distance at least $3$ along $P$, while they are connected via $q$ by just two hops, contradicting the fact that $P$ is the shortest path. 
 
To summarize, we have shown that for any node $q \in FS' \backslash P$, we have that $N(q) \cap P \leq 3$. We have also proved that in the (new) optimal \FS, $v_{k-2}$ and $v_{k-1}$ are connected to a node $u$ that has a direct link to $d$. Note that $u \in FS' \backslash P$ and it is already connected to three nodes in $P$. Hence, in the (new) optimal FS, we have that every node $v_i \in P$, $i < k-2$ must find an alternate path to at least one of the nodes $v_{k-2}, v_{k-1}, v_{k}$ or $q$ and cannot have a direct (alternate) edge to any of these nodes. We create a new graph $G''$ by collapsing the four vertices $(v_{k-2}, v_{k-1}, v_{k}, q)$ (call this new node $d'$); to compute an optimal FS on $G'$, we can compute an optimal FS on $G''$ and combine it with the edges between these nodes that form a part of $FS'$. Hence, we have reduced our problem to a strictly smaller subproblem with the same constraints. This allows us to use a simple recursion. Let $S(n)$ denote the edge-set size of the optimal FS in $G'$ with $|P| = n$. Then, we get the following recursion:
 \begin{eqnarray*}
 S(n) & \geq & S(n-2) + 5 
 \end{eqnarray*}
 which gives us the claimed lower bound on the edge-set size of the FS. To prove that the bound is indeed tight, we use the graph shown in Fig.~\ref{fig:lbu}.
 \end{proof}

\end{document}